\documentclass[preprint]{aastex}

\usepackage{amsmath}	% Package used for the split environment.
\usepackage{graphicx,color}

\usepackage{natbib}

\newcommand{\rhoi}{\rho_i}
\newcommand{\rhon}{\rho_n}
\newcommand{\pie}{p_{ie}}
\newcommand{\pn}{p_n}
\newcommand{\vi}{v_i}
\newcommand{\vn}{v_n}

\newcommand{\Ti}{T_i}
\newcommand{\Tn}{T_n}

\newcommand{\kB}{k_B}

\newcommand{\alphaen}{\alpha_{en}}
\newcommand{\alphain}{\alpha_{in}}

\newcommand{\Sigmain}{\Sigma_{in}}

\newcommand{\mi}{m_i}
\newcommand{\mn}{m_n}
\newcommand{\mprot}{m_p}

\newcommand{\rhoiO}{\rho_{i0}}
\newcommand{\rhonO}{\rho_{n0}}

\newcommand{\pieO}{p_{ie0}}
\newcommand{\pnO}{p_{n0}}

\newcommand{\Hi}{H_i}
\newcommand{\Hn}{H_n}
\newcommand{\TO}{T_0}

\newcommand{\rhobi}{\rho_{bi}}

\newcommand{\rhobn}{\rho_{bn}}

\newcommand{\rhobiO}{\rho_{bi0}}
\newcommand{\rhobnO}{\rho_{bn0}}

\newcommand{\zO}{z_0}

\newcommand{\vmax}{v_{max}}

\shorttitle{Dynamics of falling partially ionized plasma blobs in the solar atmosphere}
\shortauthors{Oliver et al.}

\begin{document}

\title{DYNAMICS OF CORONAL RAIN AND DESCENDING PLASMA BLOBS IN SOLAR PROMINENCES: II.~PARTIALLY IONIZED CASE}

\author{R. Oliver\altaffilmark{1,2}, R. Soler\altaffilmark{1,2}, J. Terradas\altaffilmark{1,2}, T. V. Zaqarashvili\altaffilmark{3,4}}
%, M. L. Khodachenko\altaffilmark{5}}
\altaffiltext{1}{Departament de F\'\i sica, Universitat de les Illes Balears, E-07122 Palma de Mallorca, Spain}
\altaffiltext{2}{Institute of Applied Computing \& Community Code (IAC$^3$), UIB, Spain}
\altaffiltext{3}{Institute of Physics, IGAM, University of Graz, Universit\"atsplatz 5, 8010, Graz, Austria}
\altaffiltext{4}{Abastumani Astrophysical Observatory, Ilia State University, 0162 Tbilisi, Georgia}
%\altaffiltext{5}{Space Research Institute, Austrian Academy of Sciences, Schmiedlstrasse 6, 8042, Graz, Austria}
\email{e-mail: ramon.oliver@uib.es}

%\author{R. Oliver and R. Soler and J. Terradas}
%\affil{Departament de F\'\i sica, Universitat de les Illes Balears, E-07122 Palma de Mallorca, Spain \\ Institute of Applied Computing \& Community Code (IAC$^3$), UIB, Spain}
%\email{e-mail: ramon.oliver@uib.es}
%
%\author{T. V. Zaqarashvili}
%\affil{Institute of Physics, IGAM, University of Graz, Universit\"atsplatz 5, 8010, Graz, Austria \\
%Abastumani Astrophysical Observatory, Ilia State University, 0162 Tbilisi, Georgia}
%
%\author{M. L. Khodachenko}
%\affil{Space Research Institute, Austrian Academy of Sciences, Schmiedlstrasse 6, 8042, \\ Graz, Austria}

\begin{abstract}
Coronal rain clumps and prominence knots are dense condensations with chromospheric to transition region temperatures that fall down in the much hotter corona. Their typical speeds are in the range 30--150~km~s$^{-1}$ and of the order of 10--30~km~s$^{-1}$, respectively, i.e., they are considerably smaller than free fall velocities. These cold blobs contain a mixture of ionized and neutral material that must be dynamically coupled in order to fall together, as observed. We investigate this coupling by means of hydrodynamic simulations in which the coupling arises from the friction between ions and neutrals. The numerical simulations presented here are an extension of those of \citet{oliver2014} to the partially ionized case. We find that, although the relative drift speed between the two species is smaller than 1~m~s$^{-1}$ at the blob center, it is sufficient to produce the forces required to strongly couple charged particles and neutrals. The ionization degree has no discernible effect on the main results of our previous work for a fully ionized plasma: the condensation has an initial acceleration phase followed by a period with roughly constant velocity and, in addition, the maximum descending speed is clearly correlated with the ratio of initial blob to environment density.
\end{abstract}


\section{INTRODUCTION}

Several relevant facts converge in the physics of coronal rain; some of them are outlined next:

\begin{itemize}
\item First, its genesis. Coronal rain blobs are presumably formed when heating at the feet of active region loops trigger catastrophic cooling events \citep{muller2004,muller2005,antolin2010b,fang2015,moschou2015}. Plasma at coronal heights then cools from coronal to chromospheric temperatures \citep{vashalomidze2015,antolin2015} and starts to fall under the action of gravity. The study of coronal rain thus has the potential of understanding coronal heating processes through their associated catastrophic cooling \citep[see][]{antolin2010b}.
\item Second, its dynamics. During their fall, rain blobs usually reach smaller than free-fall velocities and rather constant accelerations \citep{wiik1996,schrijver2001,degroof2004,degroof2005,zhang2009,antolin2010b,antolin2011,antolin2012a,antolin2012b}. The simple numerical simulations of \citet[ hereafter Paper~I]{oliver2014} indicate that the presence of the condensation in the coronal loop results in a rearrangement of the hot plasma and that a large pressure gradient builds up. This provides and upward force that allows the blob to fall at a reduced, smaller than gravity acceleration. \citet{antolin2015} give another explanation, by which the condensing material drags and compresses the magnetic field. As a consequence, the magnetic pressure downstream of the blob increases and so an upward force that counteracts gravity is established.
\item Third, the potential of coronal rain as a magnetic field tracer. \citet{antolin2011,antolin2012a,harra2014,scullion2014} have used coronal rain blobs to reveal the multi-stranded magnetic structure of the magnetic tubes in which the cool clumps move. These magnetic tubes are often part of the diffuse (and hard to resolve) component of the corona, that in a single active region contains of the order of 100,000 strands \citep{klimchuk2015}. Furthermore, \citet{antolin2012a} found that coronal rain condensations often occur simultaneously in neighboring strands, which implies that these strands have a coherent cooling. This evidence points to a heating mechanism that affects adjacent strands at the same time and in a similar manner; this brings us back to the first issue, namely the coronal rain formation.
\item Fourth, its role in the chromosphere-corona mass cycle. \citet{antolin2012a,antolin2015} estimated the downward mass flux per loop to be of the order of 1--5$\times 10^9$ g s$^{-1}$. This is a lower limit to the true mass drainage by coronal rain since the smallest mass clumps have widths below the observational resolution \citep{antolin2012a,scullion2014,antolin2015}; the same conclusion has been reached by \citet{fang2013,fang2015,moschou2015} on the basis of coronal rain numerical simulations.  According to \citet{antolin2012a}, who used the mass flux estimates of \citet{beckers1972,pneuman1978}, the mass supplied by spicules to the solar corona amounts to $1.5\times 10^{10}$ g s$^{-1}$. In addition, a condensation/drainage rate of $10^{10}$ g s$^{-1}$ during the formation of a prominence was estimated by \citet{liu2012}. It then seems that coronal rain contributes significantly to the mass interchange between chromosphere and corona.
%Multi-thermal: \citet{schrijver2001,scullion2014,antolin2015}
\item Fifth, coronal rain encompasses transition region to chromospheric temperatures and displays neutral and ionized species falling in unison. This co-spatiality of neutral and ionized material has been observed in H$\alpha$ and \ion{He}{2}~304~\AA\ \citep{degroof2004,degroof2005}, in H$\alpha$ and \ion{Ca}{2} 8542~\AA\ \citep{ahn2014}, and in H$\alpha$ and \ion{Ca}{2}~H \citep{chae2010}. The last observation corresponds to cool emission features in a hedgerow prominence. More observations, with even better spatial resolution, are desirable to have more information on the dynamics of neutrals and ions in falling plasmas.
\end{itemize}

In this work we propose to use coronal rain and falling prominence knots as a test-bed for exploring the interaction of the ionized and neutral plasma fractions. The numerical simulations of Paper~I show that the descending cool clumps tend to achieve a more or less constant speed whose value depends, among other factors, on the ratio of blob to environment (i.e., loop) density. In the absence of ion-neutral interactions, the neutral fraction of a blob falls in a loop evacuated of neutrals, so that the density ratio is very large and the falling speed becomes much larger than that of the ionized fraction of the blob. In other words, very quickly the neutral part of the cool clump would move away from the ionized part, which contradicts the observations. Therefore, ion-neutral friction must be large enough to force a common dynamics of the two species. To investigate the ion-neutral coupling, a two-fluid model is adopted here (Section~\ref{sect_model}). The coronal rain path is assumed vertical since loop curvature does not induce qualitative changes in the blob dynamics (Paper~I). A relevant assumption in our model is that once the cool blob is formed the cooling process triggered by thermal instability ceases to operate. This may contradict the observations of \citet{antolin2015}, who find that coronal rain appears intermittent and clumpy at coronal heights, although it becomes more continuous at the chromospheric level. These authors suggest that this is caused by the smearing effect of a continued cooling of the rain even after it has formed. The results and discussion are presented in Sections~\ref{sect_results} and \ref{sect_discussion}, respectively. We note that our conclusions can also be relevant for the dynamics of descending prominence knots \citep{engvold1976,liu2012}.

\section{MODEL}
\label{sect_model}

%The purpose of this work is to investigate the dynamics of gas condensations in coronal or prominence environments and to estimate the importance of partial ionization effects. We concentrate in the dynamics of the falling material after it has condensed and, therefore, we do not reproduce the condensation process, nor other physical phenomena such as the recombination of protons and electrons that takes place as gas temperature and density change. Our aim is to understand the small acceleration observed in vertical prominence threads and at evaluating the relevance of ion-neutral collisions in this process.

\subsection{Governing equations}

To describe the temporal evolution of a partially ionized blob falling under the influence of gravity together with the pressure and friction forces we consider the two-fluid equations for the charged (ions plus electrons) and neutral fractions \citep[e.g.,][]{draine1983,smith2008,zaqarashvili2011b,meier2011,meier2012,leake2014,khomenko2014}. We thus have the mass continuity equations for charged particles and neutrals, 

\begin{equation}\label{eqmassi}
\frac{\partial\rhoi}{\partial t}+\nabla\cdot(\rhoi{\bf v}_i)=0,
\end{equation}

\begin{equation}\label{eqmassn}
\frac{\partial\rhon}{\partial t}+\nabla\cdot(\rhon{\bf v}_n)=0,
\end{equation}

\noindent where $\rhoi$ and $\rhon$ are their respective densities (the electron mass is neglected) and ${\bf v}_i$ and ${\bf v}_n$ are their respective velocities. We also have the momentum equations of charged particles and neutrals,

\begin{equation}\label{eqmomi}
\rhoi\left[\frac{\partial{\bf v}_i}{\partial t}+({\bf v}_i\cdot\nabla){\bf v}_i\right]=-\nabla\pie+\rhoi{\bf g}-\alphain({\bf v}_i-{\bf v}_n),
\end{equation}

\begin{equation}\label{eqmomn}
\rhon\left[\frac{\partial{\bf v}_n}{\partial t}+({\bf v}_n\cdot\nabla){\bf v}_n\right]=-\nabla\pn+\rhon{\bf g}+\alphain({\bf v}_i-{\bf v}_n),
\end{equation}

\noindent with $\bf g$ the acceleration of gravity, $\pie$ the combined pressure of ions and electrons, and $\pn$ the neutrals pressure. Furthermore, the friction between the two plasma fractions causes their coupling through a force per unit volume proportional to ${\bf v}_i-{\bf v}_n$. The proportionality constant, $\alphain$, is written as \citep{leake2012,soler2015}

\begin{equation}\label{eq_alphain}
\alphain=\frac{\Sigmain}{\mi+\mn}\sqrt{\frac{8\kB}{\pi}\left(\frac{\Ti}{\mi}+\frac{\Tn}{\mn}\right)}\rhoi\rhon.
\end{equation}

\noindent In this expression $\kB$ is Boltzmann's constant, $\mi$ and $\mn$ are the respective atomic masses of ions and neutrals, $\Ti$ and $\Tn$ are their respective temperatures, and $\Ti/\mi+\Tn/\mn$ is their atomic mass-averaged temperature: see \citet[][ p. 47]{mitchner1973} or \citet{draine1986}. Here we assume a H plasma and so $\mn\simeq\mi=\mprot$ (with $\mprot$ the proton mass). Furthermore, the mass-averaged temperature is $(\Ti+\Tn)/2$. Hence, Equation~(\ref{eq_alphain}) is equivalent to Equation (7.4) of \citet{braginskii1965}, who assumed an equal temperature of ions and neutrals that is here replaced by this mass-averaged value. Note that since $\alphain$ depends on the ions and neutrals density and temperature, it implicitly depends on position and time. Finally, $\Sigmain$ is the collision cross-section between ions and neutrals \citep[see, e.g.,][ for its definition]{braginskii1965,chapman1970}. Slightly different values for $\Sigmain$ are given in the literature \citep[see][ for a discussion of this topic]{soler2015}. To compute this parameter, \citet{braginskii1965,zaqarashvili2011a} use the hard-spere model for collision cross-sections \citep[see][]{chapman1970} and obtain $\Sigmain=4.70\times10^{-20}$~m$^2$. \citet{leake2012} use the value $\Sigmain=1.41\times10^{-19}$~m$^2$, which is exactly three times the one used by the previous authors. On the other hand, \citet{vranjes2013} use quantum-mechanical theory to compute cross-sections of elastic scattering and charge transfer processes. From their results the value $\Sigmain\simeq 10^{-18}$ m$^2$ can be inferred. This is the value used in this work, which is in close agreement with the one used by \citet{leake2013} \citep[based on the work of][]{draine1983}, namely $\Sigmain=1.16\times 10^{-18}$~m$^2$.

We finally have the energy equations of charged particles and neutrals, written in terms of their pressures,

\begin{equation}\label{eqpressie}
%\frac{\partial\pie}{\partial t}+({\bf v}_i\cdot\nabla)\pie+\gamma\pie\nabla\cdot{\bf v}_i=(\gamma-1)\alphain({\bf v}_i-{\bf v}_n)\cdot{\bf v}_i,
\frac{\partial\pie}{\partial t}+({\bf v}_i\cdot\nabla)\pie+\gamma\pie\nabla\cdot{\bf v}_i=(\gamma-1)Q_i^{in},
\end{equation}

\begin{equation}\label{eqpressn}
%\frac{\partial\pn}{\partial t}+({\bf v}_n\cdot\nabla)\pn+\gamma\pn\nabla\cdot{\bf v}_n=-(\gamma-1)\alphain({\bf v}_i-{\bf v}_n)\cdot{\bf v}_n,
\frac{\partial\pn}{\partial t}+({\bf v}_n\cdot\nabla)\pn+\gamma\pn\nabla\cdot{\bf v}_n=(\gamma-1)Q_n^{in}.
\end{equation}

\noindent $Q_i^{in}$ and $Q_n^{in}$ represent the heat generation due to ion-neutral collisions and thermal transfer \citep{draine1986,meier2011,leake2014} and are given by

\begin{equation}\label{eqqiin}
Q_i^{in}=\alphain\left[\frac{1}{2}|{\bf v}_i-{\bf v}_n|^2+3\frac{\kB}{\mprot}(\Tn-\Ti)\right],
\end{equation}

\begin{equation}\label{eqqnin}
Q_n^{in}=\alphain\left[\frac{1}{2}|{\bf v}_i-{\bf v}_n|^2+3\frac{\kB}{\mprot}(\Ti-\Tn)\right].
\end{equation}

\noindent We note that although some authors \citep[e.g., ][]{smith2008,zaqarashvili2011b} have used different expressions for these quantities, the ones provided here are correct. It is also worth to mention that \citet{terradas2015} solved Equations~(\ref{eqmassi})--(\ref{eqqnin}) to study the support of neutrals against gravity in prominences. Although they state that $Q_i^{in}=-Q_n^{in}$ (or $W_{in}=-W_{ni}$ in their notation), this is just an error in the text and their solutions are completely valid.

%We supplement the above system of equations by adding expressions for the Lagrangian displacement of a charged or neutral plasma element, denoted by ${\boldsymbol\xi}_i$ and ${\boldsymbol\xi}_n$, respectively,

%\begin{equation}\label{eqdispli}
%\frac{\partial{\boldsymbol\xi}_i}{\partial t}={\bf v}_i-{\bf v}_i\cdot\nabla{\boldsymbol\xi}_i,
%\end{equation}

%\begin{equation}\label{eqdispln}
%\frac{\partial{\boldsymbol\xi}_n}{\partial t}={\bf v}_n-{\bf v}_n\cdot\nabla{\boldsymbol\xi}_n.
%\end{equation}

%\noindent If a charged particle is initially at position ${\bf r}$, at time $t$ its position is ${\bf r}+{\boldsymbol\xi}_i$ (and similarly for neutrals). Thus, the Lagrangian displacement allows us to characterize the motion of individual charged/neutral plasma elements.

Since our aim is to concentrate in the dynamics of a partially ionized blob falling in the solar atmosphere, a number of assumptions have been made when writing the above equations. Effects such as ionization/recombination and charge exchange interactions have been ignored. Furthermore, collisions between electrons and neutrals usually have a negligible effect and for this reason we have considered a vanishing electron-neutral friction coeficient ($\alphaen=0$) in the momentum and energy equations. Joule heating and the divergence of the heat flux in the energy equations have also been neglected. In addition, in the following only motions in the vertical direction are taken into account (see Paper~I for a discussion of the curvature of the blob path and its effect on the blob dynamics). Moreover, the effect of magnetic fields is ignored.

%A Cartesian coordinate system with the $z$-axis pointing in the vertical direction is considered. Next we insert $\rhoi=\rhoi(z,t)$, $\rhon=\rhon(z,t)$, $\pie=\pie(z,t)$, $\pn=\pn(z,t)$, ${\bf v}_i=\vi(z,t) {\bf \hat e}_z$, ${\bf v}_n=\vn(z,t) {\bf \hat e}_z$, ${\boldsymbol\xi}_i=\xii(z,t) {\bf \hat e}_z$, and ${\boldsymbol\xi}_n=\xin(z,t) {\bf \hat e}_z$ into Equations~(\ref{eqmassi})--(\ref{eqmomn}), (\ref{eqpressie}), (\ref{eqpressn}), (\ref{eqdispli}), and (\ref{eqdispln}) and substitute ${\bf g}=-g{\bf \hat e}_z$. The horizontal components of the momentum equations and of the expressions for the Lagrangian displacements are identically zero and so we end up with eight non-linear partial differential equations (PDEs) for the eight unknowns,

A Cartesian coordinate system with the $z$-axis pointing in the vertical direction is considered. Next we insert $\rhoi=\rhoi(z,t)$, $\rhon=\rhon(z,t)$, $\pie=\pie(z,t)$, $\pn=\pn(z,t)$, ${\bf v}_i=\vi(z,t) {\bf \hat e}_z$, and ${\bf v}_n=\vn(z,t) {\bf \hat e}_z$ into Equations~(\ref{eqmassi})--(\ref{eqmomn}), (\ref{eqpressie}), and (\ref{eqpressn}) and substitute ${\bf g}=-g{\bf \hat e}_z$. The horizontal components of the momentum equations are identically zero and so we end up with six non-linear partial differential equations (PDEs) for the six unknowns,

\begin{equation}\label{eqrhoi}
\frac{\partial\rhoi}{\partial t}=-\vi\frac{\partial\rhoi}{\partial z}-\rhoi\frac{\partial\vi}{\partial z},
\end{equation}

\begin{equation}\label{eqrhon}
\frac{\partial\rhon}{\partial t}=-\vn\frac{\partial\rhon}{\partial z}-\rhon\frac{\partial\vn}{\partial z},
\end{equation}

\begin{equation}\label{eqvi}
\rhoi\frac{\partial\vi}{\partial t}=-\rhoi\vi\frac{\partial\vi}{\partial z}-\frac{\partial\pie}{\partial z}-g\rhoi-\alphain(\vi-\vn),
\end{equation}

\begin{equation}\label{eqvn}
\rhon\frac{\partial\vn}{\partial t}=-\rhon\vn\frac{\partial\vn}{\partial z}-\frac{\partial\pn}{\partial z}-g\rhon+\alphain(\vi-\vn),
\end{equation}

\begin{equation}\label{eqpie}
\frac{\partial\pie}{\partial t}=-\vi\frac{\partial\pie}{\partial z}-\gamma\pie\frac{\partial\vi}{\partial z}+(\gamma-1)Q_i^{in},
\end{equation}

\begin{equation}\label{eqpn}
\frac{\partial\pn}{\partial t}=-\vn\frac{\partial\pn}{\partial z}-\gamma\pn\frac{\partial\vn}{\partial z}+(\gamma-1)Q_n^{in}.
\end{equation}

%\begin{equation}\label{eqxii}
%\frac{\partial\xii}{\partial t}=\vi\left(1-\frac{\partial\xii}{\partial z}\right),
%\end{equation}

%\begin{equation}\label{eqxin}
%\frac{\partial\xin}{\partial t}=\vn\left(1-\frac{\partial\xin}{\partial z}\right).
%\end{equation}

These expressions are supplemented with the ideal gas law for charged particles and neutrals,

\begin{equation}\label{eq_temp}
\pie=2\rhoi R\Ti, \hspace{1cm} \pn=\rhon R\Tn,
\end{equation}

\noindent with $R=\kB/\mprot$ the ideal gas constant. The reason for the presence of the factor 2 in Equation~(\ref{eq_temp}) for charged particles is that the partial pressures of ions and electrons are summed up under the conditions of same temperature and same number density.

\subsection{Static equilibrium}

In Paper~I we considered a fully ionized blob falling in a vertically stratified atmosphere representing the coronal part of the plasma along the blob path. Coronal H is fully ionized and for this reason the neutral fraction of the blob moves in an environment devoid of neutrals. This causes numerical problems when solving the two-fluid system of PDEs presented above and for this reason we here assume that the ionized and neutral fractions of the blob move in an atmosphere in which both gas fractions are present. The amount of neutrals is kept as small as possible and so ions are predominant in the coronal part of the plasma. For a homogeneous equilibrium temperature, Equations~(\ref{eqrhoi})--(\ref{eqpn}) have the following static solution

\begin{equation}\label{sol_equili}
\pie(z,t=0)=\pieO e^{-z/\Hi}, \hspace{0.5cm} \rhoi(z,t=0)=\rhoiO e^{-z/\Hi},
\end{equation}

\begin{equation}\label{sol_equiln}
\pn(z,t=0)=\pnO e^{-z/\Hn}, \hspace{0.5cm} \rhon(z,t=0)=\rhonO e^{-z/\Hn},
\end{equation}

\noindent where the density and pressure at the coronal base ($z=0$) satisfy the ideal gas law for the charged and neutral fractions of the gas,

\begin{equation}\label{eq_pieO_pnO}
\pieO=2\rhoiO R\TO, \hspace{1cm} \pnO=\rhonO R\TO.
\end{equation}

\noindent $\TO$ is the homogenous equilibrium temperature, assumed equal for all species in the initial state. Furthermore, the charged particles and neutrals vertical scale heights ($\Hi$ and $\Hn$) depend on the equilibrium temperature as follows,

\begin{equation}\label{eq_Hi_Hn}
\Hi=2\frac{R\TO}{g}, \hspace{1cm} \Hn=\frac{R\TO}{g}.
\end{equation}

\noindent Then, the isothermal equilibrium assumption results in $\Hi=2\Hn$.

\subsection{Mass condensation}
\label{sect_mass_condensation}

At $t=0$ a dense blob composed of charged and neutral material is added to the previous static atmosphere and its temporal evolution is then investigated. Since the condensation process is ignored, a fully condensed blob is initially at rest at a certain height and is left to evolve under the action of gravity, the pressure gradient, and the friction force between ions and neutrals according to Equations~(\ref{eqrhoi})--(\ref{eqpn}). At $t=0$ the ionized and neutral parts of the blob have densities described by

\begin{equation}\label{rhoblobi}
\rhobi(z,t=0)=\rhobiO\exp\left[-\left(\frac{z-\zO}{\Delta}\right)^2\right],
\end{equation}

\begin{equation}\label{rhoblobn}
\rhobn(z,t=0)=\rhobnO\exp\left[-\left(\frac{z-\zO}{\Delta}\right)^2\right],
\end{equation}

\noindent where $\rhobiO$, $\rhobnO$ are the maximum blob densities of ions and neutrals at the initial time, $\zO$ is its initial position, and its length in the vertical direction is of the order of $2\Delta$. The maximum density value, $\rhobiO+\rhobnO$, used in this work is in agreement with the rain core density estimations of \citet{antolin2015}, that are in the range $2\times 10^{10}$--$2.5\times 10^{11}$ cm$^{-3}$. Similar values have been found in numerical simulations by \citet{muller2004,muller2005,antolin2010b,fang2015}.

\section{Results}
\label{sect_results}

%The initial density distribution is the sum of $\rhoi$ in Equation~(\ref{sol_equili}) and $\rhobi$ in Equation~(\ref{rhoblobi}) for ions and the sum of $\rhon$ in Equation~(\ref{sol_equiln}) and $\rhobn$ in Equation~(\ref{rhoblobn}) for neutrals. The initial pressure of the two gas components is described by Equations~(\ref{sol_equili}) and (\ref{sol_equiln}), so that at $t=0$ the blob is not in mechanical equilibrium and starts to fall. The initial velocity and Lagrangian displacement are zero for both ions and neutrals ($\vi=\vn=0$, $\xii=\xin=0$). The temporal evolution of the system is described by Equations~(\ref{eqrhoi})--(\ref{eqpn}); details of the numerical method used to solve this system of PDEs can be found in Paper~I.

The initial density distribution is the sum of $\rhoi$ in Equation~(\ref{sol_equili}) and $\rhobi$ in Equation~(\ref{rhoblobi}) for ions and the sum of $\rhon$ in Equation~(\ref{sol_equiln}) and $\rhobn$ in Equation~(\ref{rhoblobn}) for neutrals. The initial pressure of the two gas components is described by Equations~(\ref{sol_equili}) and (\ref{sol_equiln}), so that at $t=0$ the blob is not in mechanical equilibrium and starts to fall. The initial velocity is zero for both ions and neutrals ($\vi=\vn=0$). The temporal evolution of the system is described by Equations~(\ref{eqrhoi})--(\ref{eqpn}); details of the imposed boundary conditions and the numerical method used to solve this system of PDEs can be found in Paper~I.

Unless stated otherwise, we use the following parameter values at $t=0$: the ion and neutral densities at the coronal base are $\rhoiO=4.5\times 10^{-12}$~kg~m$^{-3}$ and $\rhonO=0.5\times 10^{-12}$~kg~m$^{-3}$. The temperature is $\TO=2\times10^6$~K, which results in the scale heights of ions and neutrals $\Hi\simeq 120$~Mm and $\Hn\simeq 60$~Mm. Furthermore, the blob has an initial length given by $\Delta=0.5$~Mm and is released at the height $\zO=50$~Mm. The only two parameters that need to be fixed are the maximum blob density of ions and neutrals: $\rhobiO$ and $\rhobnO$. The ionization degree of a coronal rain blob or a falling prominence knot is not well known and for this reason we select different values of $\rhobiO$ and $\rhobnO$. The results shown in this paper correspond to two cases: first, a blob made of equal quantities of ions and neutrals ($\rhobiO=\rhobnO$; Sections~\ref{sect_blob_dynamics_no_friction} to \ref{sect_blob_kinematics}) and, second, a blob containing 10\% ions and 90\% neutrals ($\rhobnO=9\rhobiO$; Section~\ref{sect_percentage}).

% Use the figure* environment to span two columns of text. Only makes a difference when using emulateapj.
\begin{figure*}[ht!]
  \centerline{
    \scriptsize{(a)}\hspace{-10ex}
    \includegraphics[width=0.35\textwidth,angle=-90]{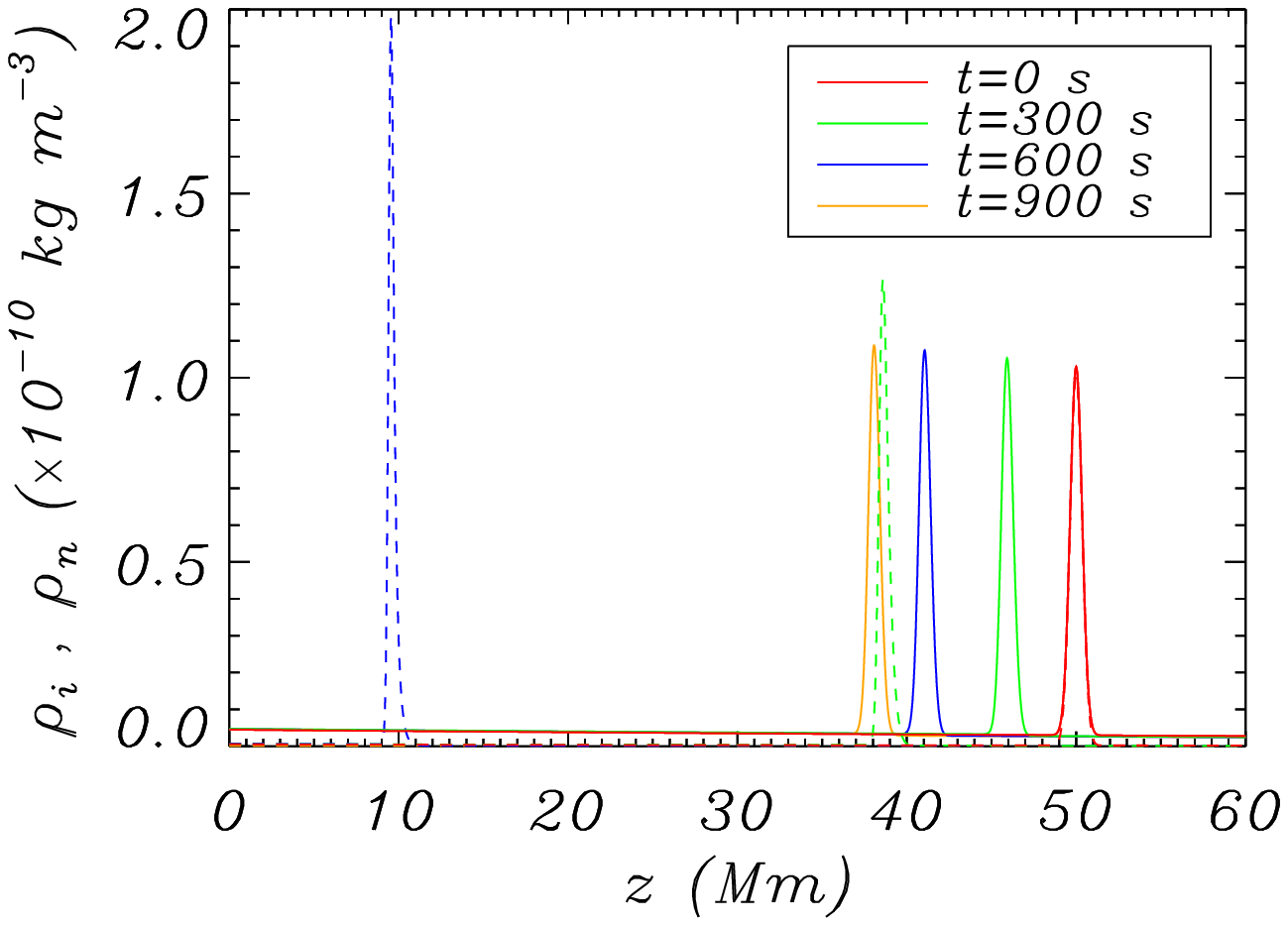} \\
    \scriptsize{(b)}\hspace{-10ex}
    \includegraphics[width=0.35\textwidth,angle=-90]{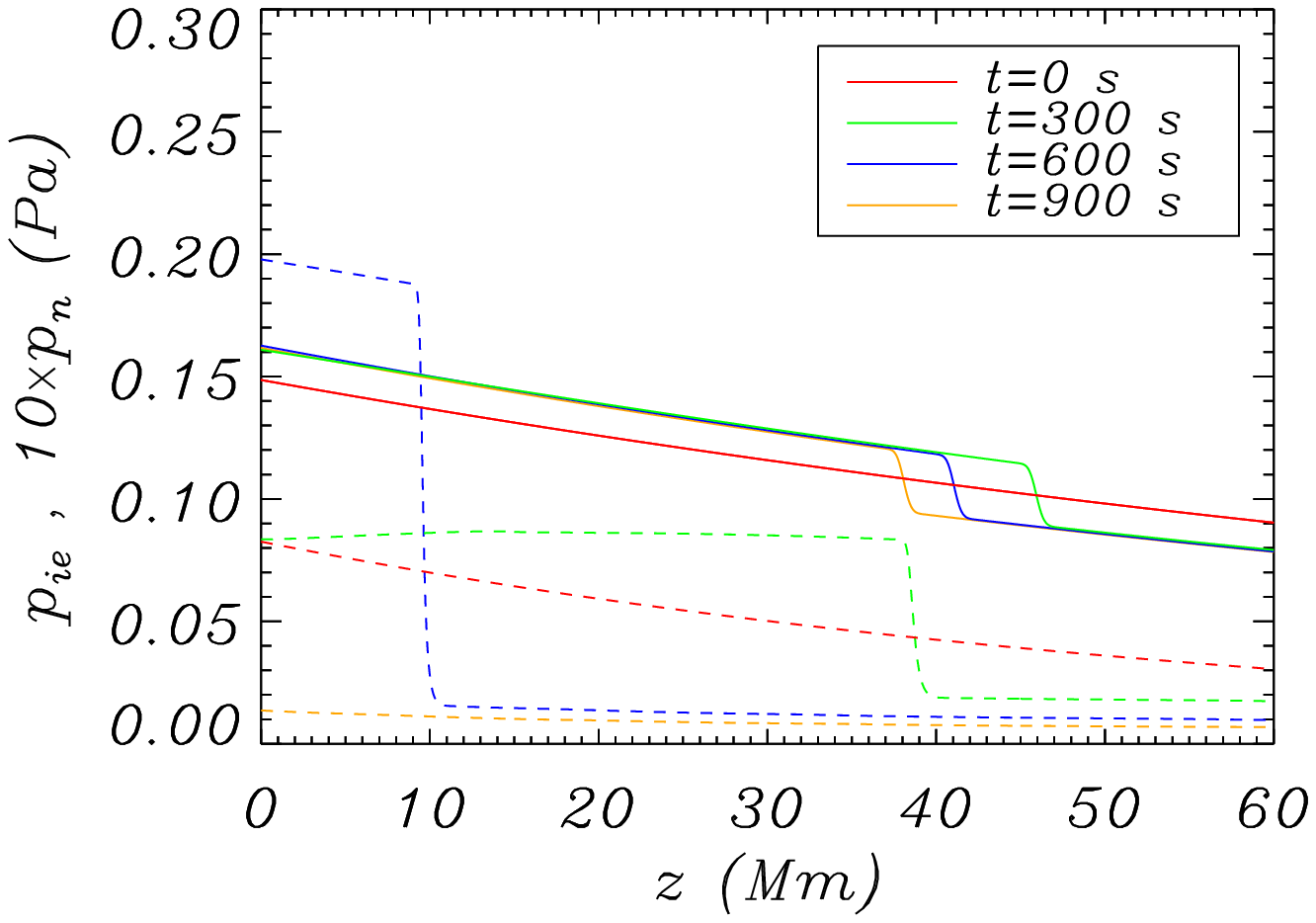} \\
  }
  \centerline{
    \scriptsize{(c)}\hspace{-10ex}
    \includegraphics[width=0.35\textwidth,angle=-90]{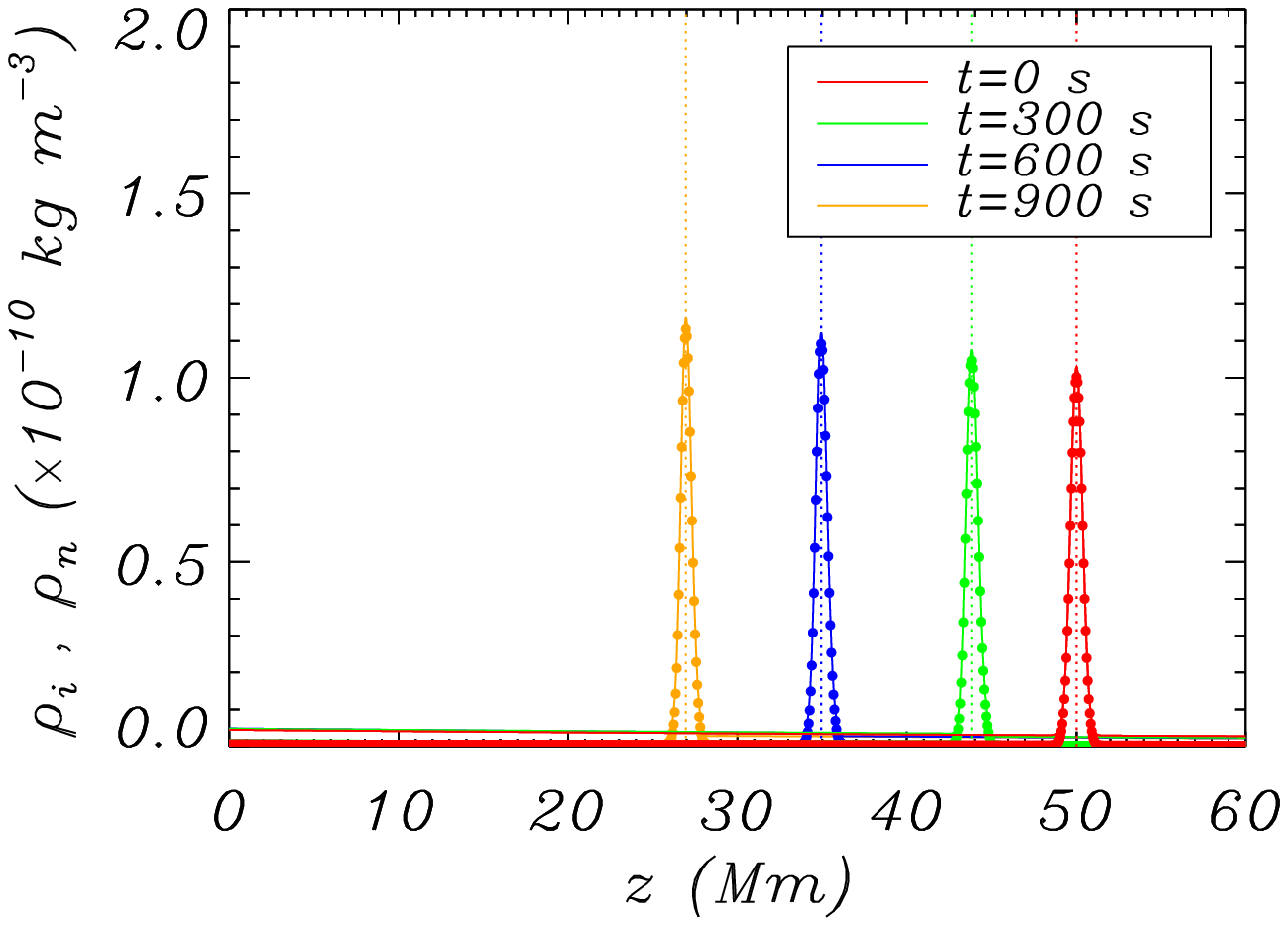} \\
    \scriptsize{(d)}\hspace{-10ex}
    \includegraphics[width=0.35\textwidth,angle=-90]{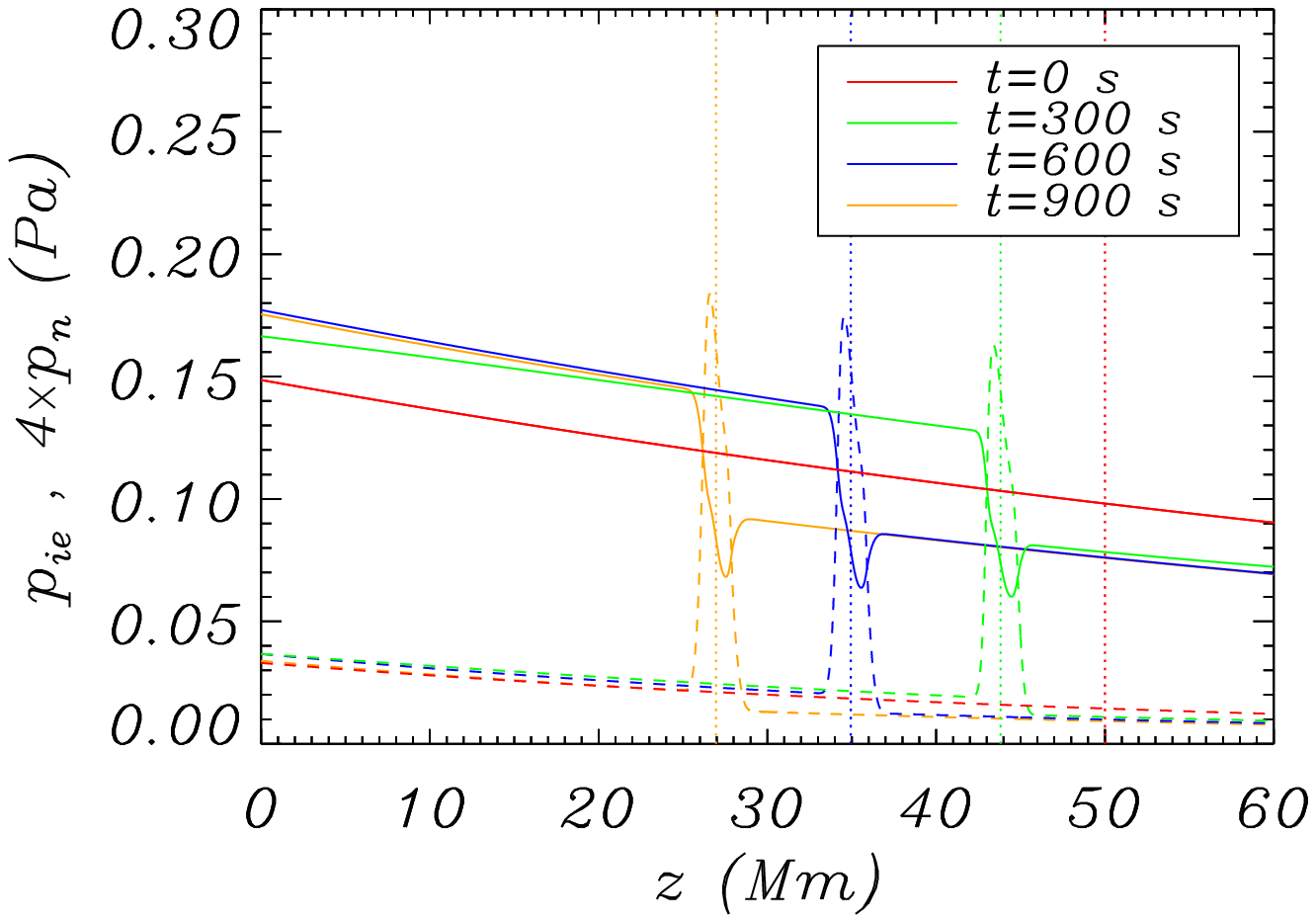}
  }
  \caption{Vertical distribution of (a) and (c) density, (b) and (d) pressure for several times. The solid and dashed lines correspond to charged particles and neutrals, whose density in panel (c) is shown with filled circles. The vertical dotted lines indicate the maximum density position. In (a) and (b) the two species are decoupled ($\alphain=0$), while in (c) and (d) the coupling between them is retained ($\alphain\neq 0$). These results have been obtained for a blob with the same amount of neutrals and ions, $\rhobiO=\rhobnO=10^{-10}$~kg~m$^{-3}$. The pressure of neutrals in (b) and (d) has been multiplied by the factor shown next to the vertical axis.}
  \label{fig_density_vs_time}
\end{figure*}

\subsection{Blob dynamics in the absence of ion-neutral friction}
\label{sect_blob_dynamics_no_friction}

We start considering the decoupled dynamics of neutrals and charged particles by imposing $\alphain=0$. Then our PDEs split into the system of Equations~(\ref{eqrhoi}), (\ref{eqvi}), and (\ref{eqpie}) for charged particles, and the independent system of Equations~(\ref{eqrhon}), (\ref{eqvn}), and (\ref{eqpn}) for neutrals. Each of these systems of PDEs is identical to that solved in Paper~I; thus, the dynamics of the neutral and charged parts of the falling blob is already known. In fact, although the blob is initially composed of a mixture of the two species, when $\alphain=0$ they will in general behave as two independent blobs, as we explain now. The main conclusion of Paper~I is that for a fixed blob length (i.e., a fixed $\Delta$), the blob dynamics is to a great extent determined by the density contrast of the blob with respect to the ambient plasma. For equal blob density ($\rhobiO=\rhobnO$), the density contrast of neutrals is larger than that of ions for two reasons: the smaller neutral density at the coronal base and the smaller neutral scale height. For this reason, neutrals fall much faster than the ionized material (Figure~\ref{fig_density_vs_time}(a)). Such as can be seen in this figure, the neutrals density also displays two side effects of a large density contrast, namely a strong increase with time of the blob density and a proportional reduction in its length so as to maintain the blob mass almost constant during its fall (these features were also found in Paper~I). Figure~\ref{fig_density_vs_time}(b) shows the pressure of the ionized and neutral fractions. We see that $\pie$ develops a gradient that remains constant and moves with the blob. This gradient is roughly equal to the gravity force, thus allowing the ionized blob to achieve a more or less constant downward speed (see Paper~I). On the other hand, because of the larger density contrast, the neutral blob requires a higher pressure gradient to counteract the gravity force. Figure~\ref{fig_density_vs_time}(b) shows how this pressure gradient builds up during the simulation.

\subsection{Blob dynamics with ion-neutral friction: temperature and ion-neutral collision frequency}
\label{sect_temp_freq}

\begin{figure*}[ht!]
  \centerline{
    \scriptsize{(a)}\hspace{-10ex}
    \includegraphics[width=0.35\textwidth,angle=-90]{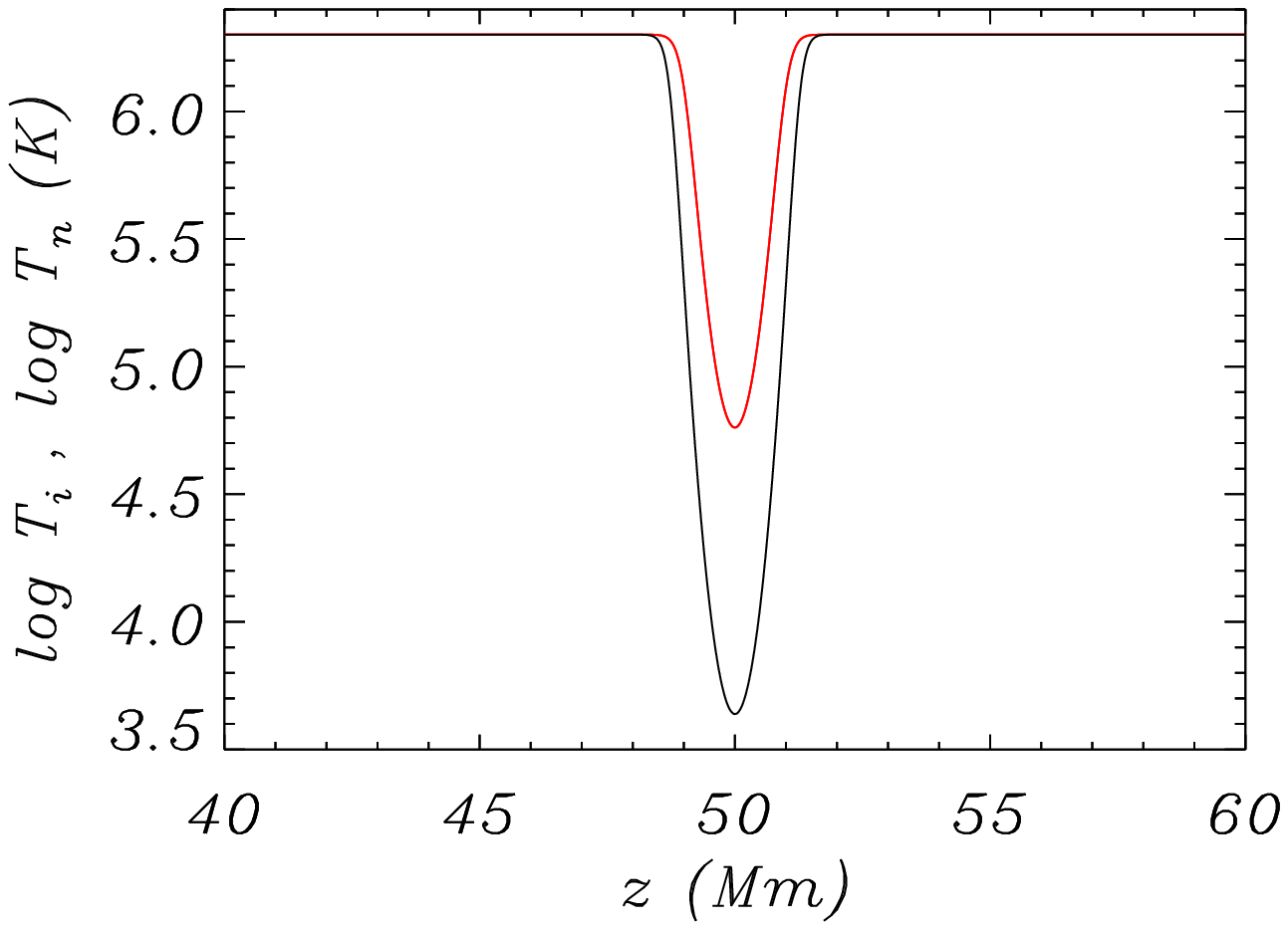} \\
    \scriptsize{(b)}\hspace{-10ex}
    \includegraphics[width=0.35\textwidth,angle=-90]{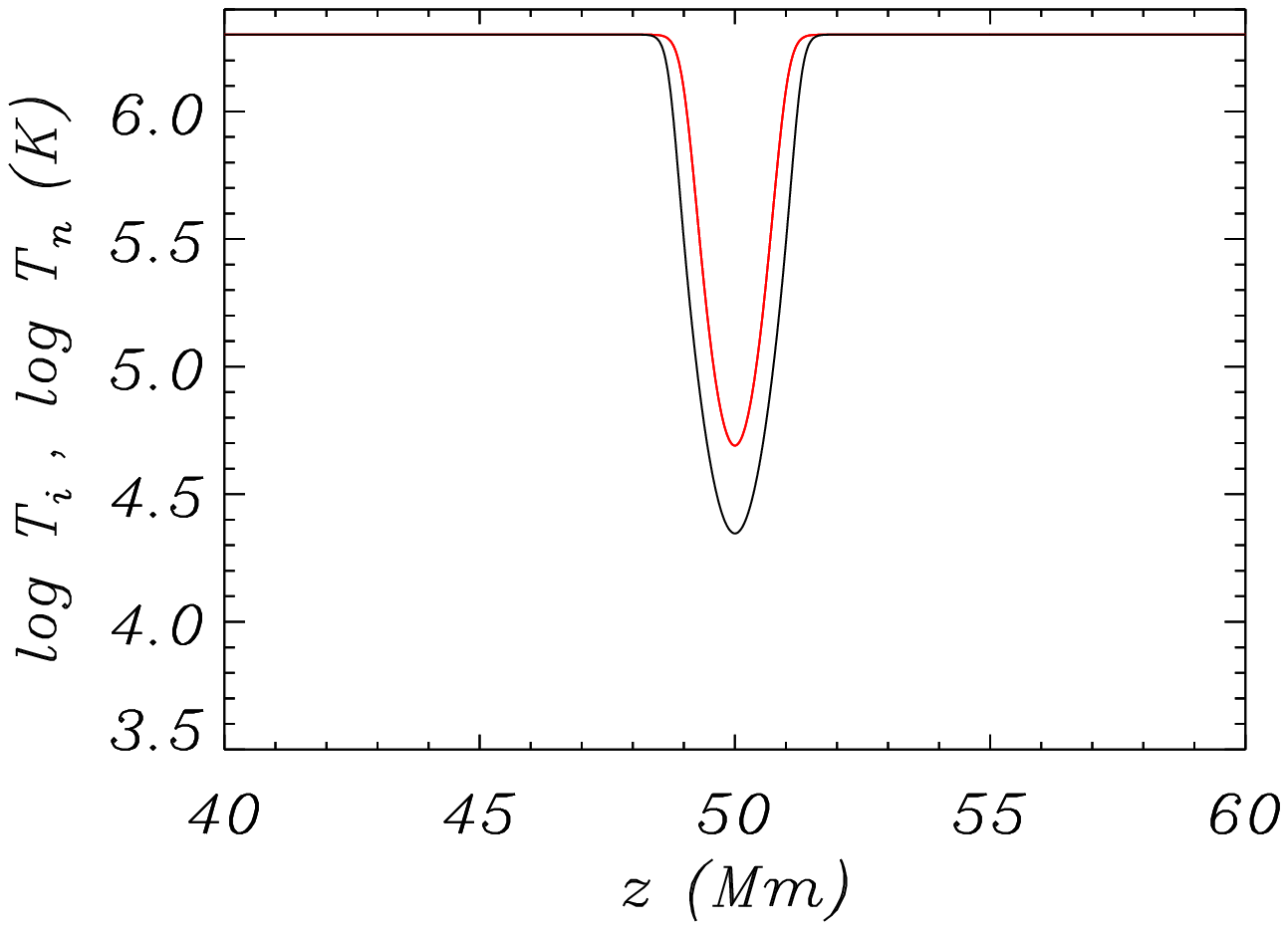} \\
  }
  \centerline{
    \scriptsize{(c)}\hspace{-10ex}
    \includegraphics[width=0.35\textwidth,angle=-90]{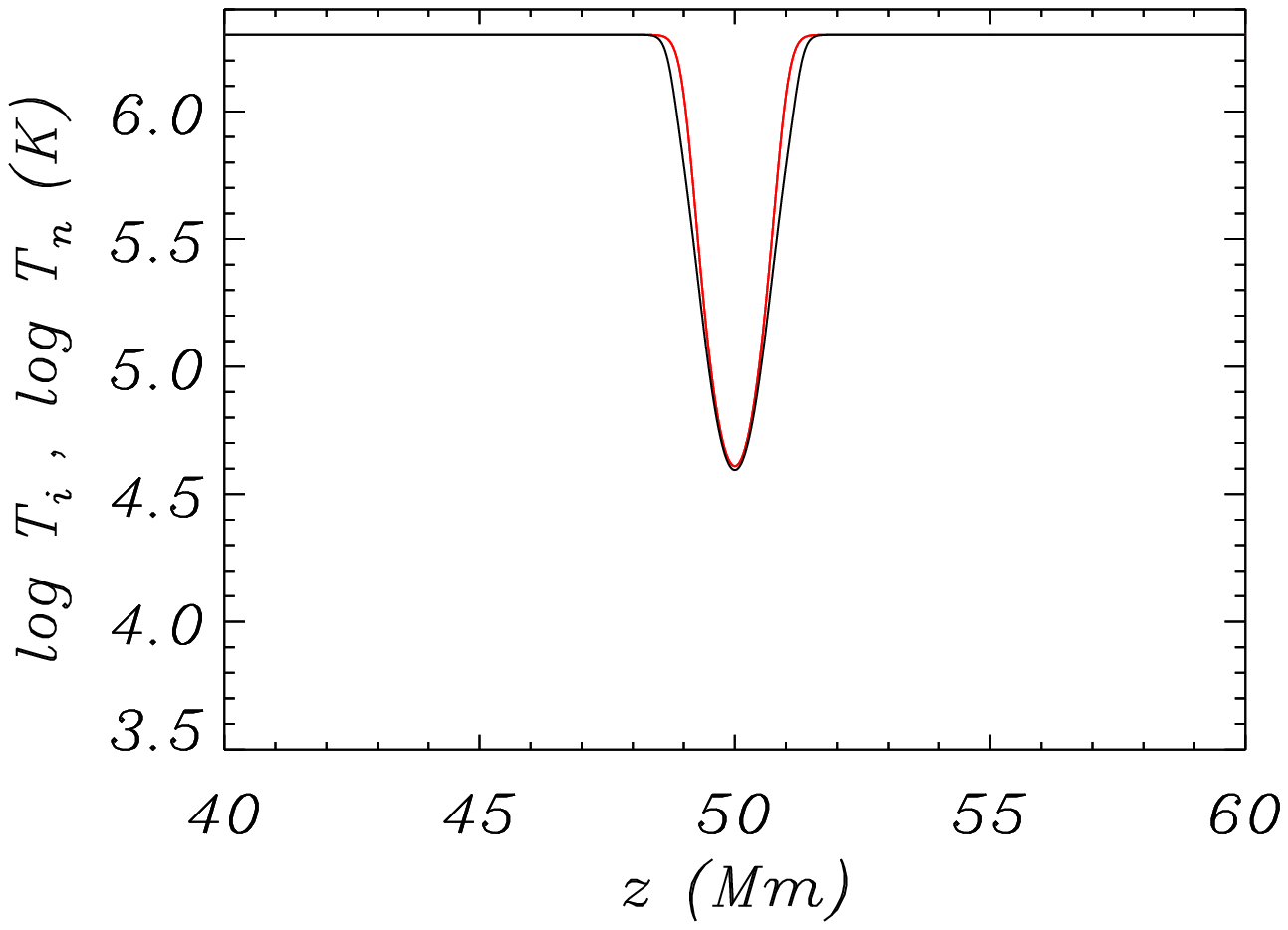} \\
    \scriptsize{(d)}\hspace{-10ex}
    \includegraphics[width=0.35\textwidth,angle=-90]{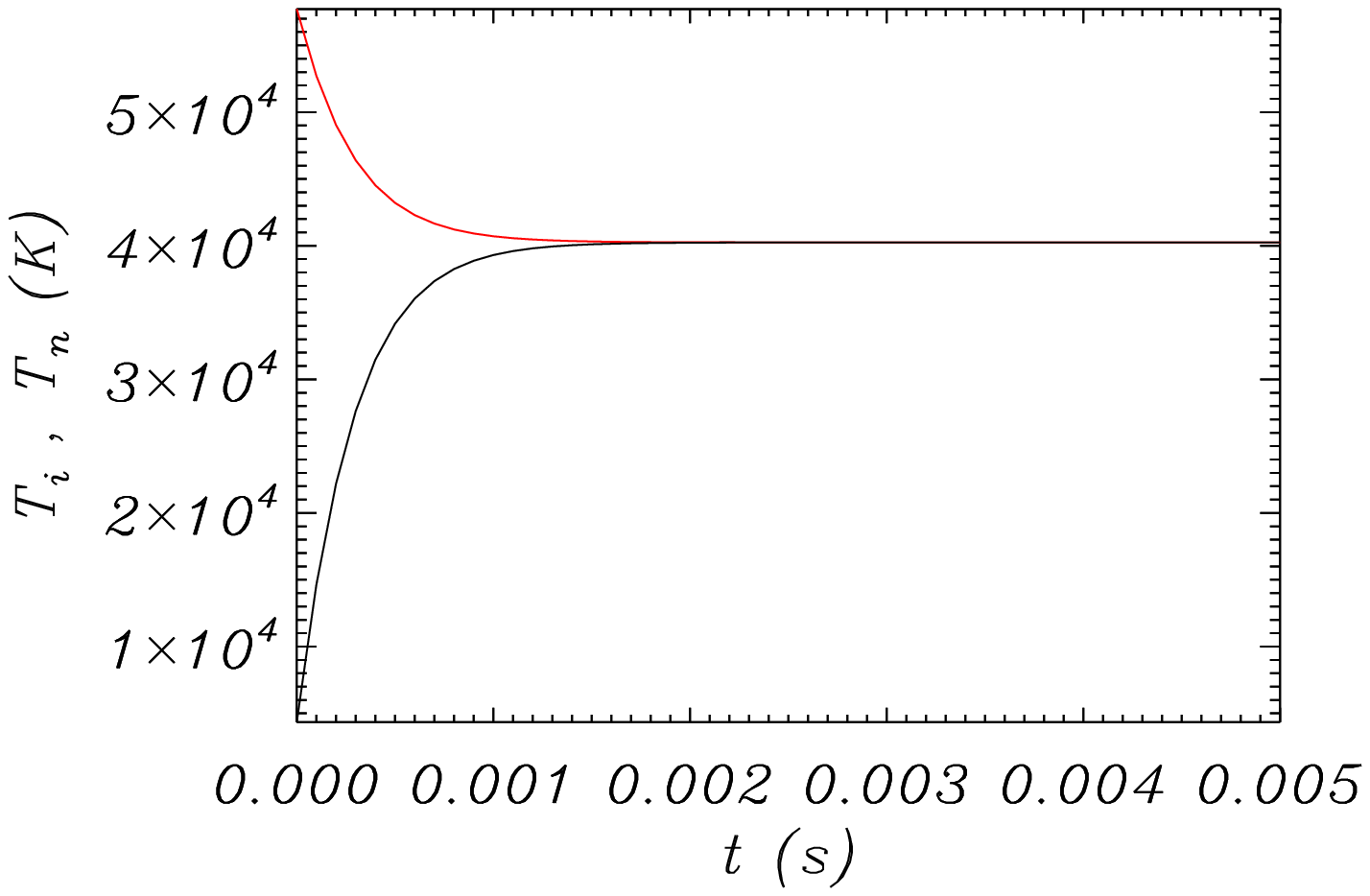}
  }
  \caption{Vertical distribution of the temperature for (a) $t=0$, (b) $t=0.2$~ms, (c) $t=1$~ms. (d) Temporal variation of the temperature at the maximum density position. Red and black lines respectively correspond to ions and neutrals. These results correspond to the numerical simulation of Figures~\ref{fig_density_vs_time}(c) and (d).}
  \label{fig_temp_vs_time}
\end{figure*}

Now we repeat the study of Section~\ref{sect_blob_dynamics_no_friction} but use the values of the friction coefficient, $\alphain$, given by Equation~(\ref{eq_alphain}). The time variation of the density and pressure for selected times are presented in Figures~\ref{fig_density_vs_time}(c) and (d). These figures are explained in Section~\ref{sect_blob_dynamics_friction} because to understand the spatial structure of the pressure we first need to pay attention to the blob temperature, that is determined from the ideal gas law for neutrals and charged particles. At $t=0$ (Figure~\ref{fig_temp_vs_time}(a)) the temperatures of charged particles and neutrals are everywhere equal to $2\times10^6$~K, except at the blob, where they smoothly decrease to their minimum values $\Ti\simeq 57700$~K and $\Tn\simeq 4350$~K. For $t>0$ (Figures~\ref{fig_temp_vs_time}(b), (c), and (d)) the two temperatures very quickly become identical, in about 2~ms, and take values around 40000~K during the whole simulation. This fast temperature balance is achieved by the thermal energy transfer induced by the difference $\Ti-\Tn$ in the last term of the quantities $Q_i^{in}$ and $Q_n^{in}$ (Equations~(\ref{eqqiin}) and (\ref{eqqnin})). 

\citet{zaqarashvili2011b} show that the relative velocity between ions and neutrals decreases exponentially in time, as $\exp(-t/\tau)$. \citet{soler2013} find an identical behavior for the vorticity component parallel to the equilibrium magnetic field. We next check if the energy interchange between ions and neutrals found here also has the same time scale, which is given by

\begin{equation}
\tau=\frac{1}{\nu_{ni}+\nu_{in}},
\end{equation}

\noindent where $\nu_{ni}$ is the collision frequency between neutrals and ions,

\begin{align}\label{eq_freq}
\nu_{ni}=\frac{\alphain}{\rhon}&=72.5\left(\frac{\Ti+\Tn}{10^4~{\rm K}}\right)^{1/2}\frac{n_i}{10^{16}~{\rm m}^{-3}} \nonumber \\
&=433\left(\frac{\Ti+\Tn}{10^4~{\rm K}}\right)^{1/2}\frac{\rhoi}{10^{-10}~{\rm kg~m}^{-3}}.
\end{align}

\noindent The collision frequency between ions and neutrals, $\nu_{in}$, follows from this formula with the subscripts $i$ and $n$ interchanged and with $\alphain=\alpha_{ni}$. Plugging in this expression the characteristic values of the blob core at $t=0$ ($\Ti\simeq 57700$~K, $\Tn\simeq 4350$~K, and $\rhoi\simeq\rhon\simeq 10^{-10}$~kg~m$^{-3}$) we obtain $\nu_{ni}\simeq\nu_{in}\simeq 1080$~Hz, i.e., a time scale $\tau\simeq 0.5$~ms. The typical time required for the temperatures $\Ti$ and $\Tn$ to equalize is of the order of a few times $\tau$, which agrees extremely well with Figure~\ref{fig_temp_vs_time}(d).

% Use the figure* environment to span two columns of text. Only makes a difference when using emulateapj.

\subsection{Blob dynamics with ion-neutral friction: density, pressure, and velocity drift}
\label{sect_blob_dynamics_friction}

From Figure~\ref{fig_density_vs_time}(c) it is obvious that ion-neutral friction provides a coupling strong enough so as to force the neutral and charged fractions of the blob to move as one. The full blob now falls as a single entity, at an intermediate speed between that of the decoupled ionized and neutral blobs of Figure~\ref{fig_density_vs_time}(a). Visual inspection of Figure~\ref{fig_density_vs_time}(c) points out that the falling speed is more or less constant; this will be discussed later.

We next turn our attention to the pressure structure (see Figure~\ref{fig_density_vs_time}(d)). Again, a pressure gradient develops at the blob position, although its spatial structure is more complex than that of the decoupled case. As we explain next, this is a consequence of the neutrals and charges temperature equalizing just described. If the thermal exchange terms in the energy equation are set to zero, then the two species are dynamically coupled through the terms proportional to ${\bf v}_i-{\bf v}_n$ in the momentum equations and the condensation falls exactly as in Figure~\ref{fig_density_vs_time}(c). Neutrals and charged particles are also coupled by means of the term proportional to $({\bf v}_i-{\bf v}_n)^2$ in the energy equations, but the omission of the quantities proportional to $\Ti-\Tn$ makes their temperatures independent from one another. The neutral and ionized parts of the blob fall together and pressure gradients similar to those in Figure~\ref{fig_density_vs_time}(b) develop. We recall that these pressure gradients result from the presence of the blob at $t=0$ and its lack of mechanical equilibrium in the vertically stratified atmosphere. On the other hand, when the terms proportional to $\Ti-\Tn$ in the energy equation are retained the ions temperature quickly varies at the beginning of the simulation from $\simeq 57700$~K to $\simeq 40000$~K and this induces a pressure decrease on top of the pressure gradient at the blob position. Hence the shape of solid lines in Figure~\ref{fig_density_vs_time}(d). In the case of neutrals, the temperature readjustment is much larger (from $\simeq 4350$~K to $\simeq 40000$~K) and so the pressure increase at the blob position prevails over the pressure gradient established at $t=0$ (dashed lines in Figure~\ref{fig_density_vs_time}(d)).

\begin{figure}[ht!]
  \scriptsize{(a)}\hspace{-10ex}
  \includegraphics[width=0.35\textwidth,angle=-90]{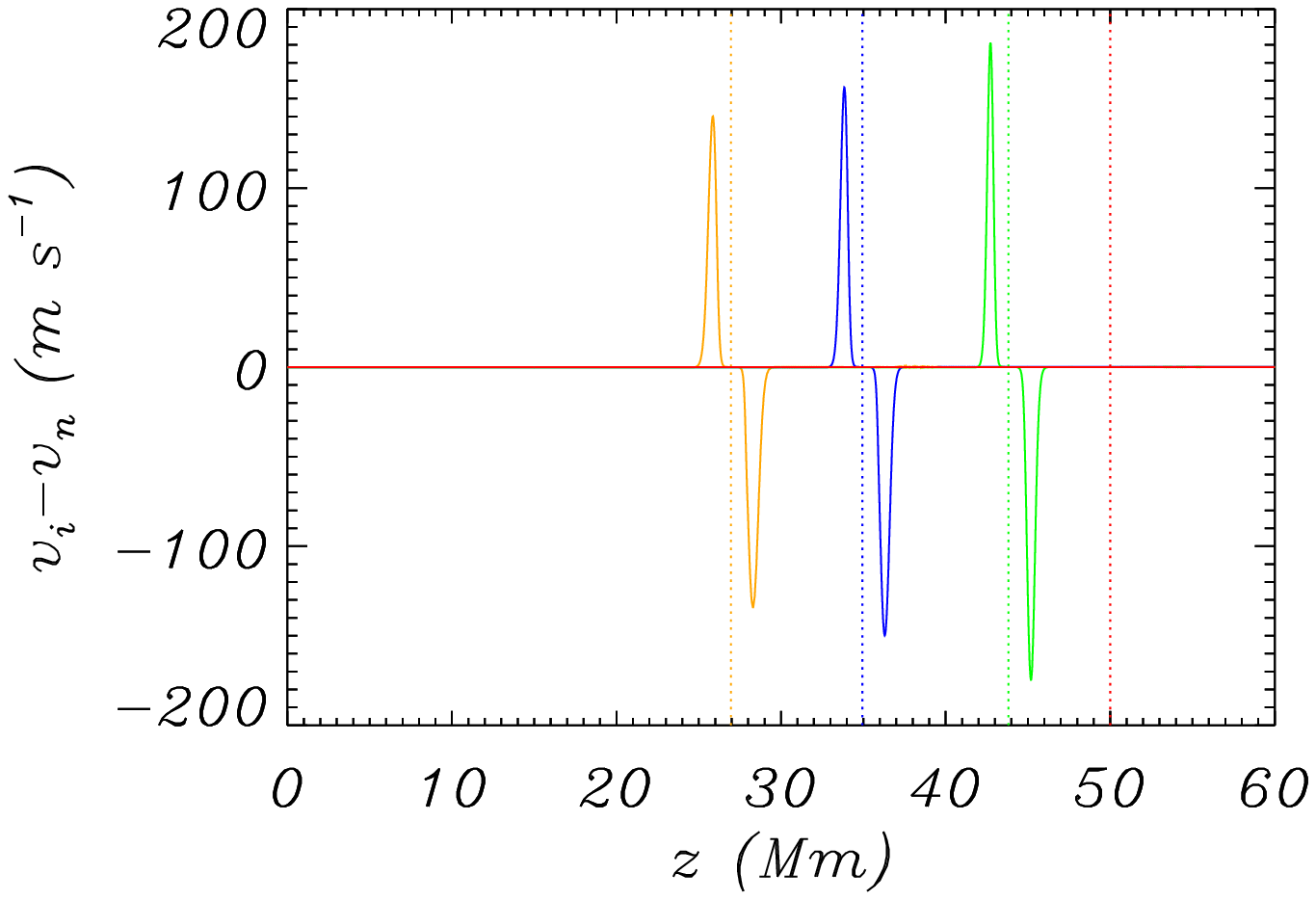} \\
  \scriptsize{(b)}\hspace{-10ex}
  \includegraphics[width=0.35\textwidth,angle=-90]{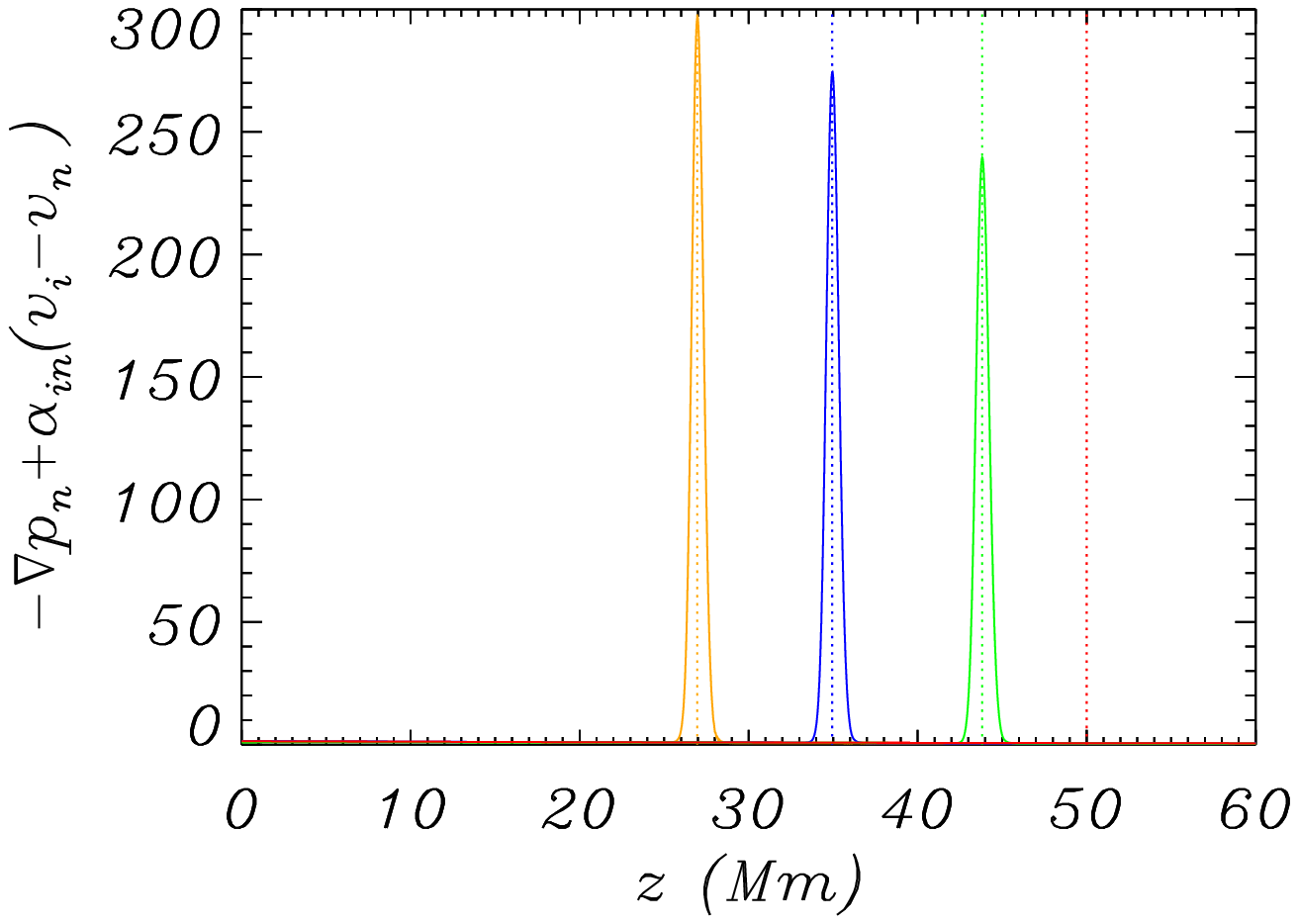}
  \caption{(a) Difference between the ion and neutral speed (in m~s$^{-1}$) and (b) sum of the pressure gradient and friction forces (per unit volume, in units of $10^{-10}$~N~m$^{-3}$) acting on neutrals, for the simulation of Figures~\ref{fig_density_vs_time}(c) and (d). The meaning of the vertical dotted lines and the line colors can be found in Figure~\ref{fig_density_vs_time}.}
  \label{fig_velocity_drift}
\end{figure}

This vertical structure of $\pie$ and $\pn$ points out a complex force balance inside the condensation. The pressure gradient at the bottom (top) part of the neutral blob, i.e., left (right) of the vertical dotted line in Figures~\ref{fig_density_vs_time}(c) and (d), points downward (upward). The opposite applies to the ionized part of the condensation and so the pressure gradient alone tends to expand the neutral fraction of the blob and to compress the ionized fraction. Nevertheless, Figure~\ref{fig_density_vs_time}(c) does not display changes in the blob shape, the reason being that the friction force acts in the opposite direction to the pressure gradient. In the case of neutrals the friction force (per unit volume) equals $\alphain(\vi-\vn)$. The vertical distribution of the velocity drift, $\vi-\vn$, for the times of Figure~\ref{fig_density_vs_time}(c) and (d) is shown in Figure~\ref{fig_velocity_drift}. One can see that it provides and upward force on neutrals at the bottom part of the blob and a downward pull at its top. These forces balance $-\nabla\pn$ and prevent the neutral blob from expanding vertically, and similarly for the ionized part of the blob, that is not compressed vertically.

It may then seem that the friction force is important in keeping the blob structure, but this is only true because of our choice of initial conditions. Such as we have mentioned, although the initial temperature difference of neutrals and charged particles inside the blob is removed in a time $\lesssim 2$~ms, it gives rise to a drift $|\vi-\vn|\simeq 200$~m~s$^{-1}$ at the blob edges. This drift is absent when the terms proportional to $\Ti-\Tn$ in the energy equations are ignored and also when the initial blob densities of neutrals and ions are selected so as to have equal initial temperatures at the blob center. Hence, the ion-neutral velocity drifts of Figure~\ref{fig_velocity_drift}(a), and the resulting complex pressure structure of Figure~\ref{fig_density_vs_time}(d), are unrealistic features of our simulations and they can be ignored because they do not modify the blob dynamics, whose study is the main aim of this paper.

The true main role of the friction force, on the other hand, is to provide the necessary forces for the common descent of ions and neutrals in the condensation. One must bear in mind that when we set $\alphain=0$ we found that neutrals fall much faster than charged particles. To see the net effect of the competing pressure gradient and friction force when friction is included, their sum is represented in Figure~\ref{fig_velocity_drift}(b). Although the drift velocity is dominant and has opposite signs on the blob edges, when these two forces are combined together they give a net upward (positive) force acting on the whole blob that makes neutrals fall more slowly than in the uncoupled case. It is worth mentioning that $\vi-\vn\simeq 0.07$~m~s$^{-1}$ at the blob center, much smaller than the value $|\vi-\vn|\simeq 200$~m~s$^{-1}$ found at the blob edges. Note that Figure~\ref{fig_velocity_drift}(b) corresponds to neutrals; when charges are considered, the sum $-\nabla\pie-\alphain(\vi-\vn)$ gives a downward (negative) force that increases the falling acceleration of charged particles. Then, the descent of neutrals is slowed down and that of ions is accelerated in the exact amount to make them fall as one.

%The blob dynamics will now be characterized attending to the behavior of the maximum density position of the condensation.

% Since the neutral and ionized fractions of the blob fall together, the position of this gradient coincides for neutrals and charged particles at any given time. Furthermore, when one compares the size of the gradient of $\pie$ in Figures~\ref{fig_density_vs_time}(b) and (d) it is clear that the upward pressure force acting on charged particles is larger when they are coupled to neutrals. The force balance required to keep a constant velocity of the charged blob fraction demands an additional downward force to compensate for the reduced pressure gradient. This force comes from the friction with neutrals. Now, Equations~(\ref{eqvi}) and (\ref{eqvn}) show that the neutrals must experience an equal upward force, so that the required pressure gradient to achieve a constant speed in Figure~\ref{fig_density_vs_time}(d) is smaller than that of Figure~\ref{fig_density_vs_time}(b). It is worth to mention that the gradient of $\pn$ in the decoupled case (Figure~\ref{fig_density_vs_time}(b)) is not enough to compensate the acceleration of gravity.

\begin{figure}[ht!]
  \scriptsize{(a)}\hspace{-10ex}
  \includegraphics[width=0.35\textwidth,angle=-90]{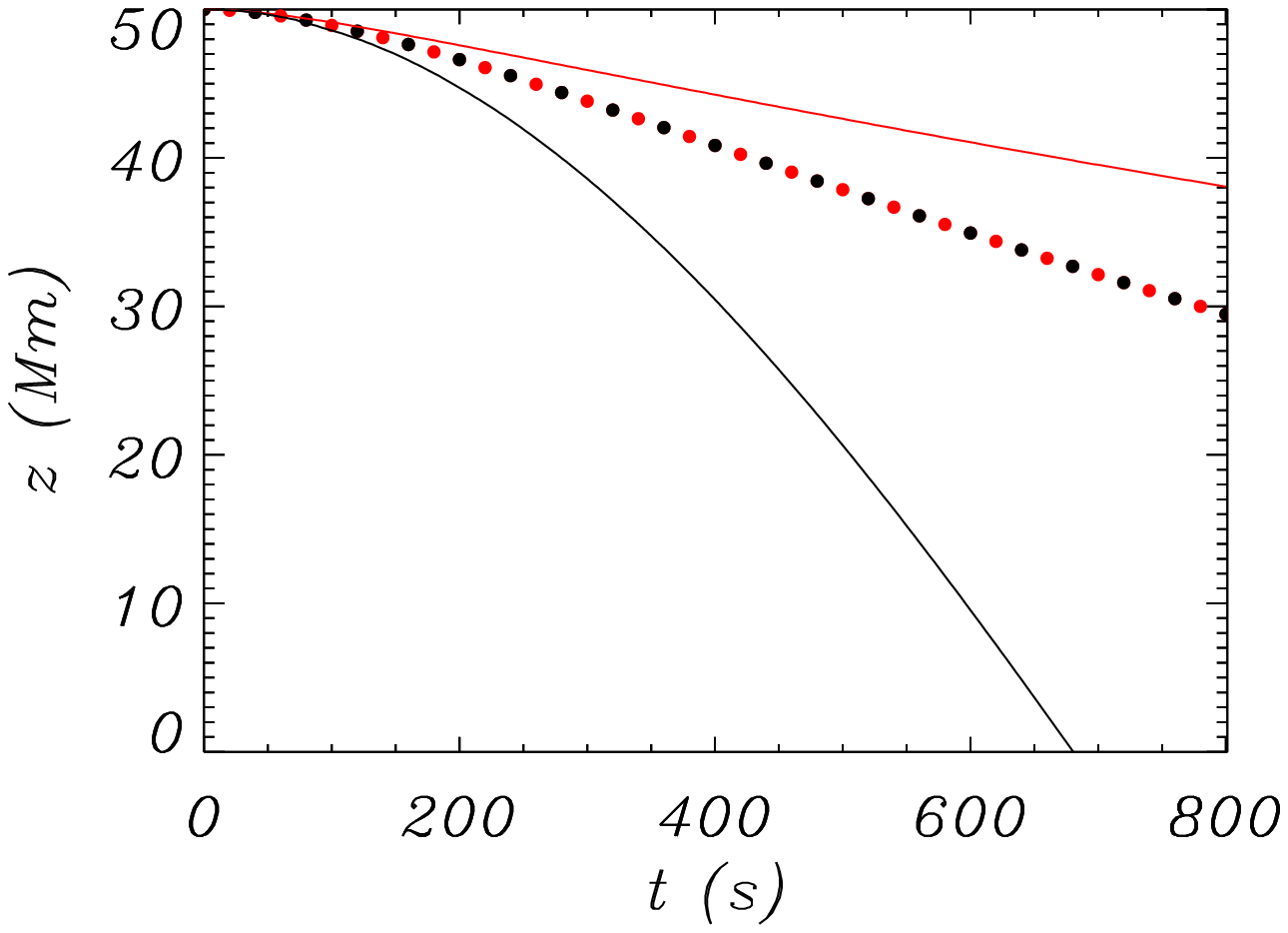} \\
  \scriptsize{(b)}\hspace{-10ex}
  \includegraphics[width=0.35\textwidth,angle=-90]{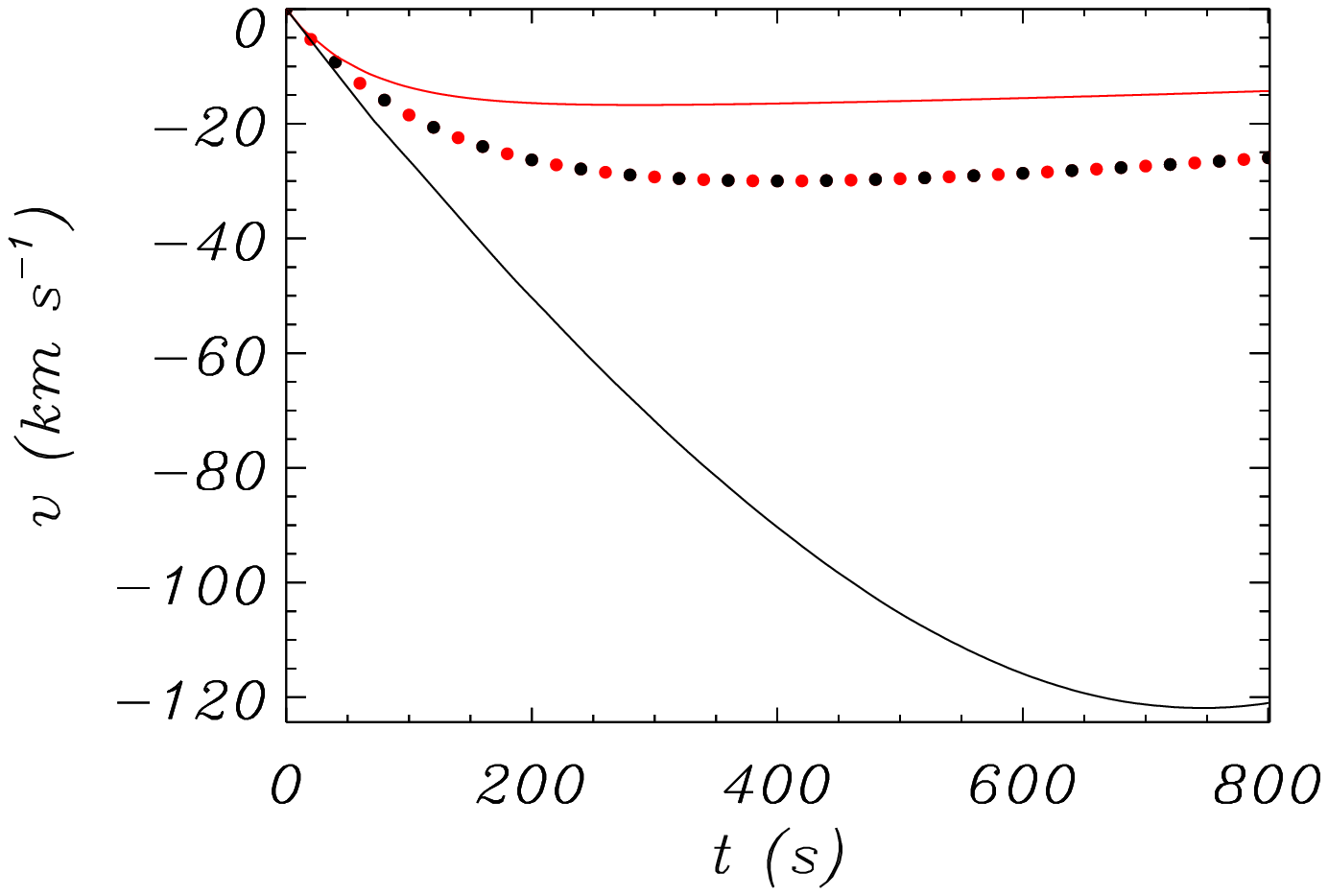}
  \caption{(a) Blob height and (b) blob velocity as a function of time for the simulations of Figure~\ref{fig_density_vs_time}. Red and black respectively correspond to ions and neutrals. The decoupled solution ($\alphain=0$) is represented by solid lines, whereas the solution in which ion-neutral friction is retained ($\alphain\neq 0$) is represented by filled circles.}
  \label{fig_no_friction_vs_friction}
\end{figure}

\subsection{Blob kinematics with ion-neutral friction}
\label{sect_blob_kinematics}

More insight into the blob kinematics can be gained by representing the height and velocity of the maximum density position, both for charges and neutrals; here we identify the behavior of the maximum density position with that of the blob. Figure~\ref{fig_no_friction_vs_friction} confirms previous results of Figure~\ref{fig_density_vs_time}: the two blob fractions move at very different speeds throughout the simulation when $\alphain=0$ (solid lines), but they move as a single entity when $\alphain\neq0$ (symbols). Perhaps the most interesting result of Figure~\ref{fig_no_friction_vs_friction} is that the falling speed in the presence of ion-neutral friction is not the average of the uncoupled values; it is actually much closer to the small velocity of the uncoupled ionized blob. The results presented here are in agreement with those of a fully ionized plasma: the blob speed increases during some time and then achieves a practically constant value, so that the height versus time diagram is initially curved followed by a more or less straight trajectory (compare Figure~\ref{fig_no_friction_vs_friction} with Figures~1(c) and (d) of Paper~I.)

\begin{figure}[ht!]
  \scriptsize{(a)}\hspace{-10ex}
  \includegraphics[width=0.35\textwidth,angle=-90]{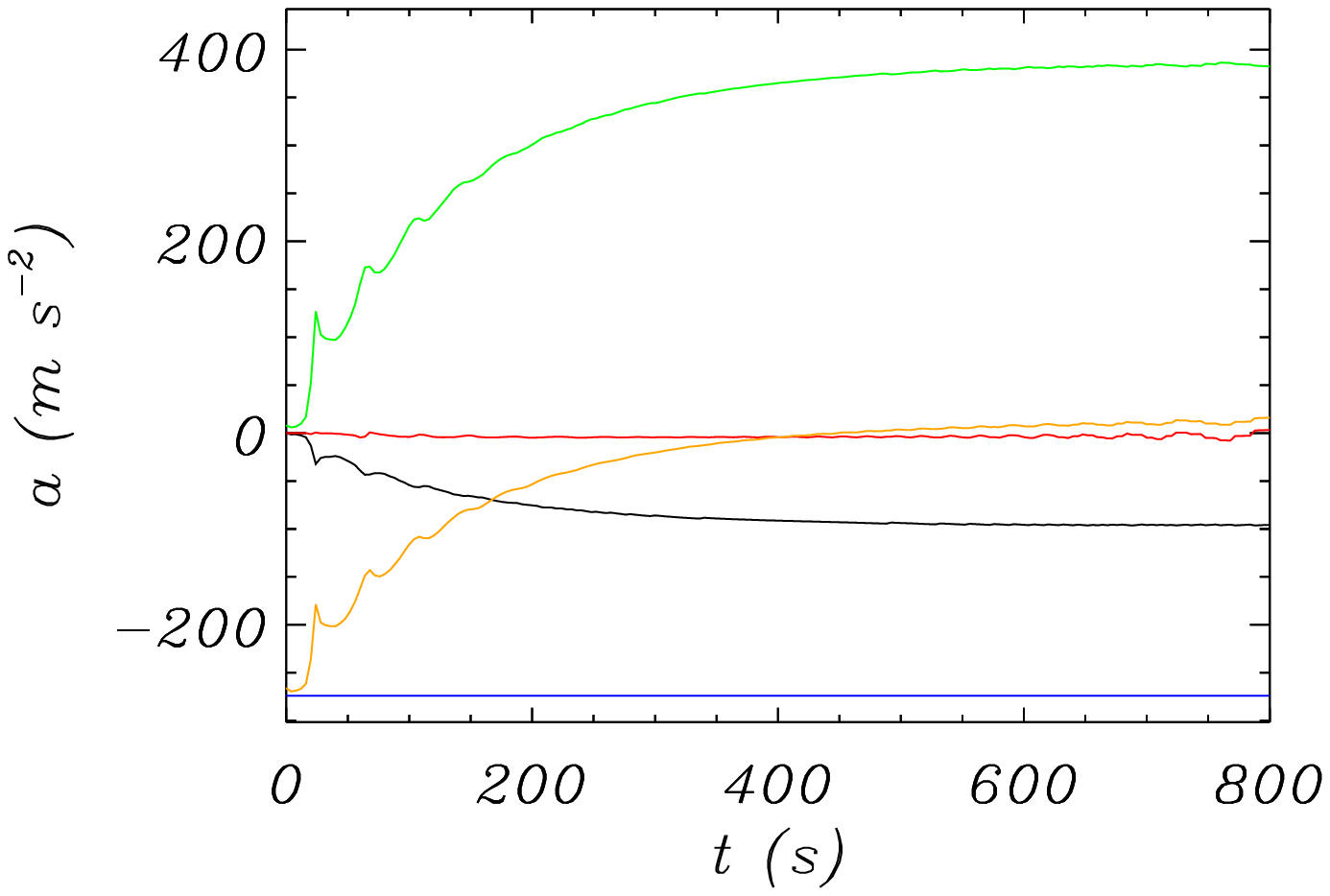} \\
  \scriptsize{(b)}\hspace{-10ex}
  \includegraphics[width=0.35\textwidth,angle=-90]{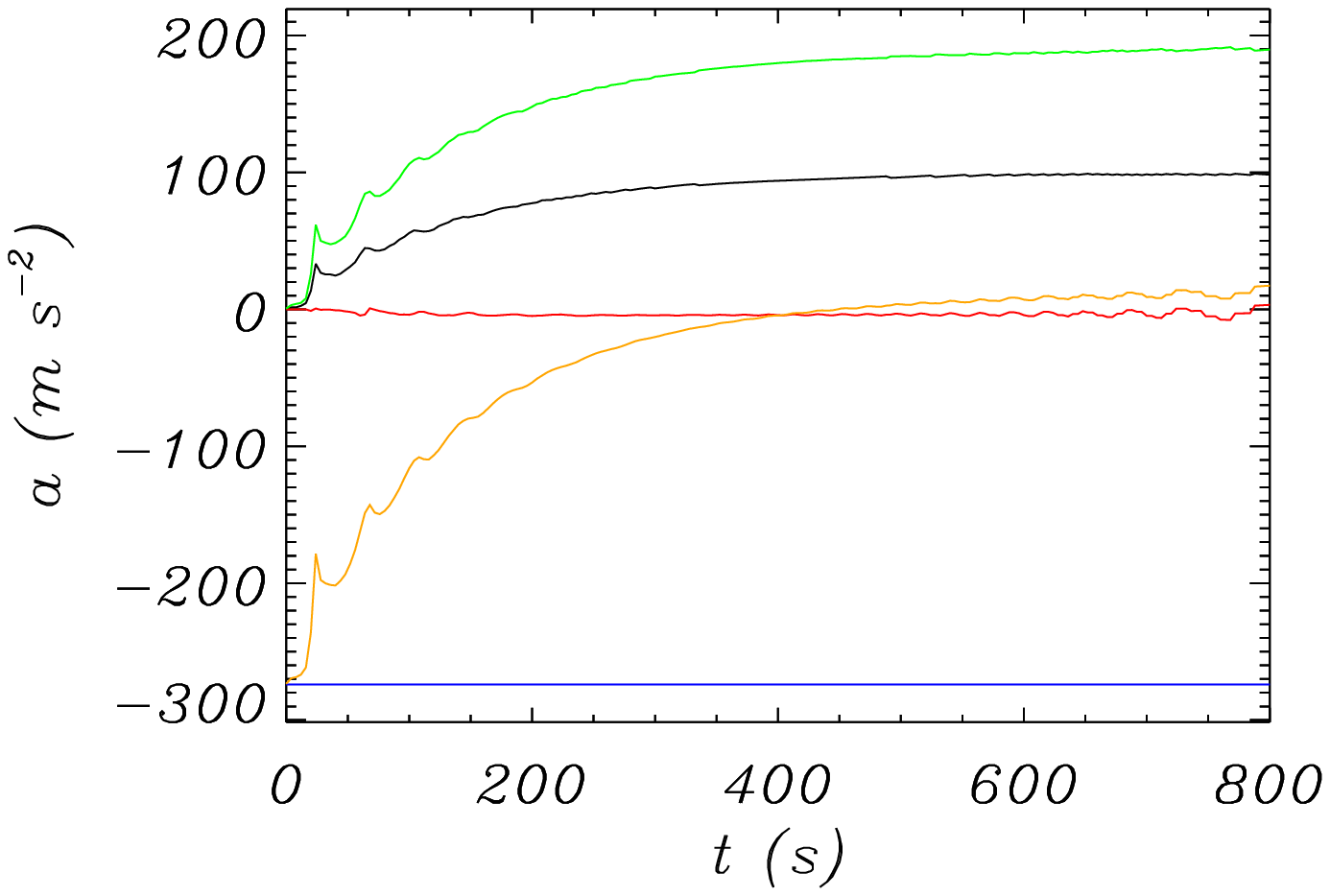}
  \caption{Blob acceleration caused by the terms in the momentum equation as a function of time for the simulations of Figures~\ref{fig_density_vs_time}(c) and (d). Red, green, blue, and black colors correspond to the inertial term, the pressure gradient, gravity, and the friction force, respectively. The total acceleration is represented by an orange line.}
  \label{fig_blob_accel}
\end{figure}

We next represent the acceleration (and the various terms contributing to it) at the maximum density position of neutrals and ions; see Figure~\ref{fig_blob_accel}. This is achieved by dividing Equations~(\ref{eqvi}) and (\ref{eqvn}) by $\rhoi$ and $\rhon$, respectively. The quantity on the left-hand side then gives us the total acceleration, while the terms on the right-hand side are the contributions to the acceleration associated to the inertial term, pressure gradient, gravity, and friction force. For both charged particles and neutrals the inertial term causes a negligible acceleration, such as corresponds to the present value of the density contrast between the blob and the environment (see Paper~I for more details). In the case of charged particles (Figure~\ref{fig_blob_accel}(a)), the upward acceleration caused by the pressure gradient gradually increases and soon overcomes that of gravity. Simultaneously, ion-neutral friction results in a downward acceleration increasing with time that, together with the acceleration of gravity, essentially lead to a force equilibrium around $t=400$~s. This is the time when the blob velocity becomes approximately constant; see filled circles in Figure~\ref{fig_no_friction_vs_friction}(d). The total acceleration in Figure~\ref{fig_blob_accel}(a) exhibits some clear oscillations at the beginning of the simulation. In Paper~I we showed that they are caused by the emission of sound waves from the blob. Next, we examine the acceleration of the neutral part of the blob (Figure~\ref{fig_blob_accel}(b)). Now the friction force points upward and reaches a value so large that together with the pressure gradient almost balances the downward pull of gravity. Given that the neutral and charged fractions of the blob fall together, the force equilibrium is also achieved at $t\simeq 400$~s. Let us also mention that the interplay of the friction and pressure forces described in this paragraph was grossly anticipated when we discussed Figure~\ref{fig_density_vs_time}(d).

The accelerations associated to the inertial term (red curves in Figure~\ref{fig_blob_accel}) display some growing oscillations at the end of the simulation whose origin is simply numerical. These oscillations cause the total blob acceleration to take values up to $\sim 20$~km~s$^{-1}$ and have no discernible consequence on the blob dynamics, e.g., the blob height and velocity in Figure~\ref{fig_no_friction_vs_friction} do not show the effect of this oscillating acceleration.

\begin{figure}[ht!]
  \scriptsize{(a)}\hspace{-10ex}
  \includegraphics[width=0.35\textwidth,angle=-90]{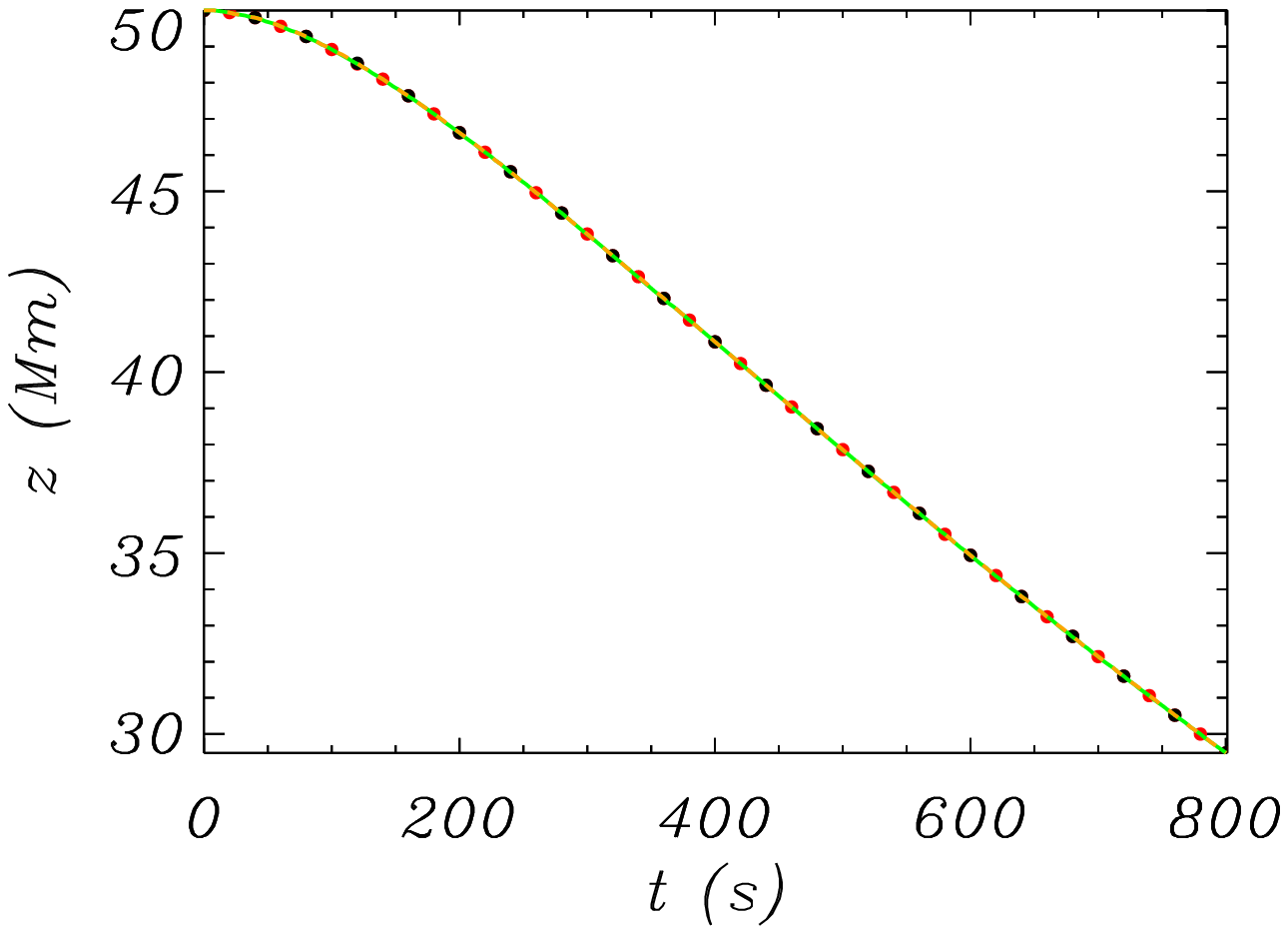} \\
  \scriptsize{(b)}\hspace{-10ex}
  \includegraphics[width=0.35\textwidth,angle=-90]{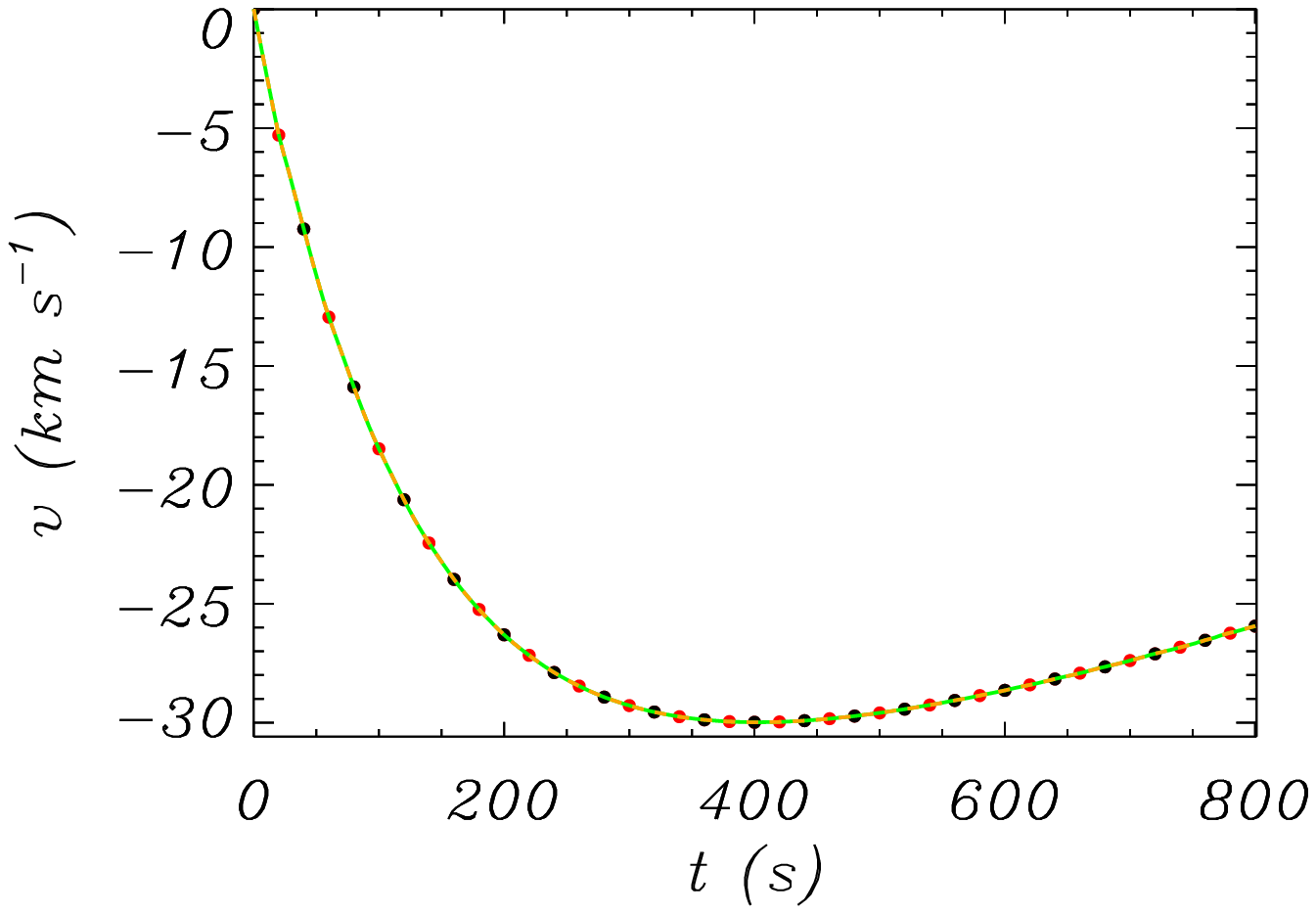}
  \caption{(a) Blob height and (b) blob velocity as a function of time for two blob ionization degrees. The filled circles have been taken from Figure~\ref{fig_no_friction_vs_friction} and correspond to a blob with equal proportion of ions and neutrals. The green (ions) and orange (neutrals) dashed lines are for a blob with the same total mass and 10\% ions and 90\% neutrals. These lines overlap and can be difficult to ascertain.}
  \label{fig_ionization_degree_z_v}
\end{figure}

\begin{figure}[ht!]
  \scriptsize{(a)}\hspace{-10ex}
  \includegraphics[width=0.35\textwidth,angle=-90]{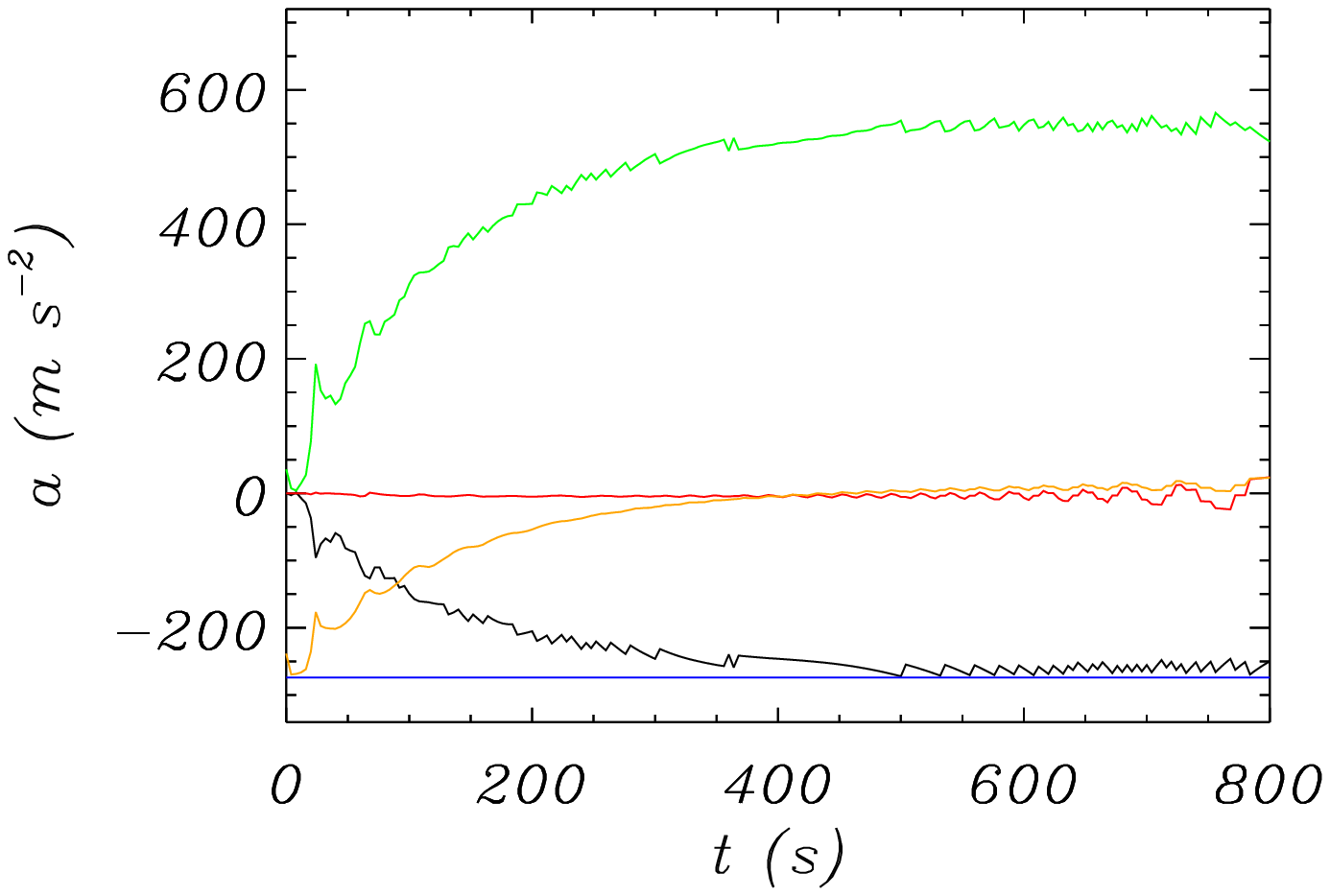} \\
  \scriptsize{(b)}\hspace{-10ex}
  \includegraphics[width=0.35\textwidth,angle=-90]{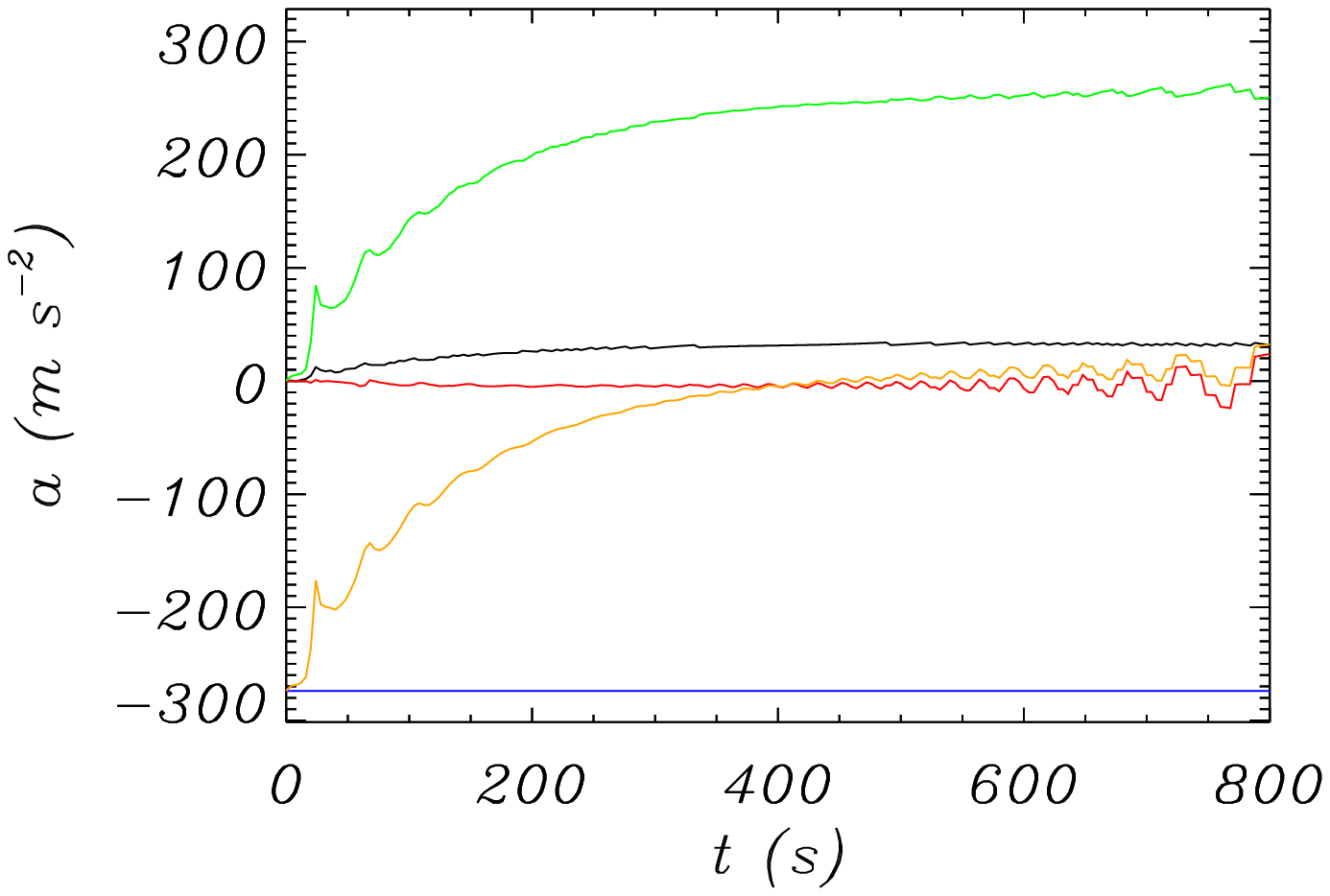}
  \caption{Same as Figure~\ref{fig_blob_accel} for a blob with 10\% ions and 90\% neutrals.}
  \label{fig_ionization_degree_a}
\end{figure}

\subsection{Effect of the ionization degree on the blob dynamics}
\label{sect_percentage}

In the results presented so far the maximum density of ions and neutrals in the blob is the same, namely $\rhobiO=\rhobnO=10^{-10}$~kg~m$^{-3}$. The relative abundance of the two species in coronal rain is controlled by ionization and recombination processes, that are ignored in this work. Therefore, our choice of $\rhobiO$ and $\rhobnO$ is arbitrary and, if incorrect, may lead to the wrong blob dynamics. Now we present the results for a blob with a very different ionization degree to ascertain the effect of this parameter. We use $\rhobiO=0.2\times 10^{-10}$~kg~m$^{-3}$, $\rhobnO=1.8\times 10^{-10}$~kg~m$^{-3}$, which implies that the total blob mass is equal to the one used before. With the new ionization fraction, the charged and neutral species also fall in unison because of the friction force. In addition, the blob height and velocity are indistinguishable from those obtained before (see Figure~\ref{fig_ionization_degree_z_v}). The same applies to the blob acceleration, which is shown in Figure~\ref{fig_ionization_degree_a} with an orange line. Other line colors refer to the various terms in the momentum equation. A comparison of the charged particles acceleration for the two ionization fractions (Figures~\ref{fig_blob_accel}(a) and \ref{fig_ionization_degree_a}(a)) shows that, although the total acceleration is the same, the acceleration caused by the pressure gradient and friction force are much larger in the blob with smaller percentage of ions. This difference is much less pronounced in the case of neutrals: see Figures~\ref{fig_blob_accel}(b) and \ref{fig_ionization_degree_a}(b). The main conclusion of this section is that the blob ionization fraction is irrelevant in the blob dynamics.

\subsection{Blob maximum falling velocity}
\label{sect_falling_speed}

\begin{figure}[ht!]
  \includegraphics[width=0.35\textwidth,angle=-90]{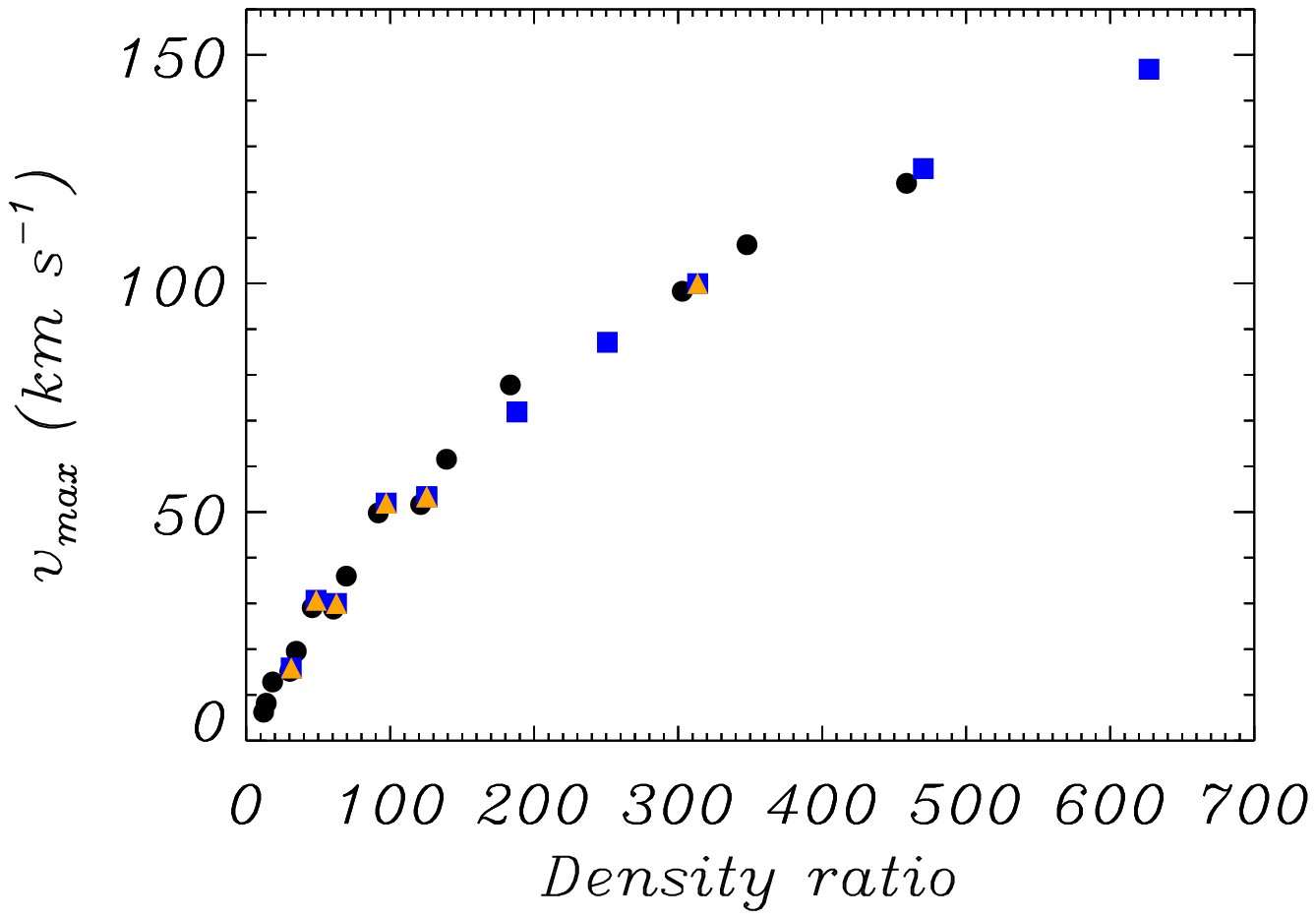}
  \caption{Maximum blob velocity (in absolute value) as a function of the initial ratio of blob to environment density. Filled circles are associated to a fully ionized plasma (Paper~I), while squares and triangles correspond to a partially ionized plasma with equal amounts of ions and neutrals and with 10\% and 90\% of ions and neutrals, respectively.}
  \label{fig_falling_speed}
\end{figure}

In Paper~I we defined the density ratio as the ratio of maximum blob density to initial coronal density at the blob position, computed from Equation~(\ref{sol_equili}) with $z$ substituted by the blob height at $t=0$. Here the density ratio is defined as the sum of the neutrals and ions density ratios, where the environment density at the initial blob position is calculated from Equations~(\ref{sol_equili}) and (\ref{sol_equiln}). In Paper~I it was found that the density ratio determined the maximum descending speed, $\vmax$, which is the maximum unsigned value of $v$; for example, in the simulation of Figure~\ref{fig_ionization_degree_z_v}(b) $\vmax\simeq30$~km~s$^{-1}$. The variation of $\vmax$ with the density ratio for a fully and a partially ionized plasma is presented in Figure~\ref{fig_falling_speed}, where the values of the filled circles are taken from Paper~I. The blue squares (orange triangles) correspond to numerical simulations in which the blob is 50\% ionized and  50\% neutral (10\% ionized and  90\% neutral) material. They follow the same trend of the fully ionized case. Moreover, we have seen before that the blob ionization degree has no effect on its dynamics, and so the position of squares and triangles coincide in this plot. Therefore, the blob maximum speed is insensitive to the blob ionization degree, even in the limit of a fully ionized blob.

\subsection{Velocity drift}
\label{sect_velocity_drift}

%The friction force only acts if there is a difference between the ion and neutral velocities at a given position. It is proportional to the velocity drift between the two species, $\vi-\vn$, where the proportionality factor, $\alphain$, depends on height and time. On the other hand, a large velocity drift can cause the ionized and neutral parts of the blob to fall separately; this is the contrary to what is observed in our simulations, so we expect $\vi-\vn$ to be small but non zero.

\begin{figure}[ht!]
  \includegraphics[width=0.35\textwidth,angle=-90]{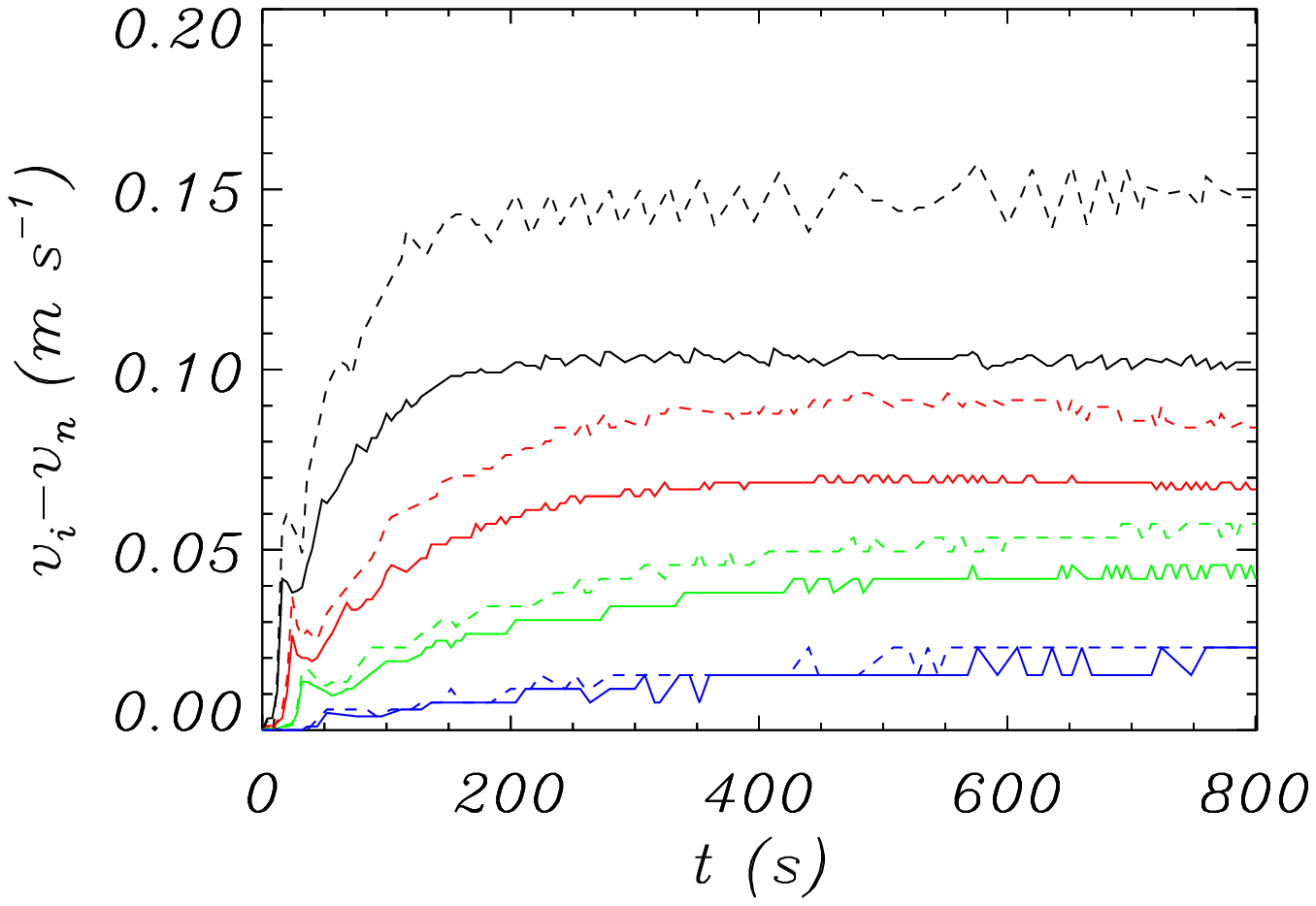}
  \caption{Difference between the ion and neutral speed (in m~s$^{-1}$) at the maximum density position as a function of time. The maximum blob density is $\rhobiO=\rhobnO=10^{-10}$~kg~m$^{-3}$ (black), $2\times 10^{-10}$~kg~m$^{-3}$ (red), $4\times 10^{-10}$~kg~m$^{-3}$ (green), and $10\times 10^{-10}$~kg~m$^{-3}$ (blue). Solid lines: the blob has equal amounts of ions and neutrals; dashed lines: the condensation contains 10\% ions and 90\% neutrals, as in Section~\ref{sect_percentage}. The red line corresponds to the simulation whose results have been presented in Figures~\ref{fig_density_vs_time} to \ref{fig_blob_accel}. The irregularities in the curves seem to be caused by insufficient precision of the numerical results because of the small velocity differences involved.}
  \label{fig_velocity_drift2}
\end{figure}

In Figure~\ref{fig_velocity_drift2} we plot the drift velocity at the position of maximum density as a function of time. The various colors and line styles represent different initial blob density and ionization degree. We obtain drift velocities (of the order of 0.01--0.15~m~s$^{-1}$) much smaller than the blob speed (of the order of 10--100~km~s$^{-1}$). Then, over the course of a few hundred seconds the neutral and ionized fractions of the blob drift by less than 1~km, an insignificant amount. Figure~\ref{fig_velocity_drift2} also shows that the drift velocity steadily increases in time until it reaches a more or less constant value. For the initial densities $\rhobiO=\rhobnO=10^{-10}$~kg~m$^{-3}$ (red solid line), the time required for the drift velocity to reach this value is of the order of 400~s, after which the blob acceleration approximately vanishes, such as shown in Figures~\ref{fig_no_friction_vs_friction}(b) and \ref{fig_blob_accel}. When the initial blob peak density is increased (solid lines) the time needed for $\vi-\vn$ to estabilise also increases because it takes longer to achieve a negligible blob acceleration; more details on this issue are given in Paper~I. We also remark that the sign of $\vi-\vn$ is positive and so neutrals fall faster than charged particles (in agreement with the uncoupled case of Figure~\ref{fig_no_friction_vs_friction}). Hence, the former provide a downward friction force on the later, as discussed in Section~\ref{sect_blob_dynamics_friction}.

Finally, if the maximum blob density is kept but its ionization degree is varied (compare solid and dashed lines with same color) the drift velocity suffers an increase that becomes larger for denser blobs. Such an increase in $\vi-\vn$ helps provide an additional upward force on neutrals so that they can fall together with the ions present in the blob.

%Figure~\ref{fig_velocity_drift2}(a) corresponds to a blob containing the same proportions of ions and neutrals. If we now consider a blob made of 10\% ionized material and 90\% neutral gas, we obtain the drift velocity presented in Figure~\ref{fig_velocity_drift2}(b), that displays all the temporal features we have just described. The main discrepancy between panels (a) and (b) lies in the value of the drift velocity, which is a factor 2 larger in the second case. Such an increase in $\vi-\vn$ helps provide an additional upward force on neutrals so that they can fall together with the ions present in the blob.

\section{Discussion}
\label{sect_discussion}

In this work we have studied the dynamics of a partially ionized coronal rain blob or falling prominence knot. We have extended the model of Paper~I (for the fully ionized case) and have included separate mass continuity, momentum, and energy equations for the charged and neutral fractions of the plasma. These equations contain the frictional interaction between neutrals and ions, and so the coupling between the two species is only dynamic (through the friction coefficient $\alphain$), while ionization/recombination and charge exchange interactions have been neglected. This means that the relative density of ions and neutrals in the blob is arbitrarily imposed. For this reason we have considered different ionization degrees of the blob. Our results show that this parameter is irrelevant and that the mass of the cold clump determines its dynamics. Hence, any physical mechanism that does not change the clump's mass as it falls but varies its ionization degree is not expected to modify the falling blob behavior.

In our simple model we have also set aside some important physics. We have excluded plasma cooling and heating, together with thermal conduction. Hence, our model does not self-consistently lead to the formation of cold coronal rain condensations through a catastrophic cooling process and so we are forced to include a plasma clump as a localized density enhancement at $t=0$. This clump then finds itself in a rather quiescent environment compared to that in which a catastrophic cooling event has just taken place. Moreover, coronal rain blobs often increase their length during their fall, which is not observed in our simulations. On the contrary, both in Paper~I and in this work falling clumps tend to become shorter in time, an effect that is more perceptible for denser clumps. This can perhaps be the consequence of omitting some physical ingredients in the energy equations.

The ion-neutral friction force is large enough to make the ionized and neutral fractions of the cool clump fall together. This friction force is proportional to the ion-neutral drift velocity, which for typical coronal rain and prominence knot conditions is of the order of 0.01--0.15~m~s$^{-1}$ at the center of the condensation. This is an insignificant value compared to the typical descending velocities of 10--100~km~s$^{-1}$. The main forces acting on the condensation are gravity, the pressure gradient (caused by the rearrangement of the pressure in the blob surroundings), and the friction force (that is proportional to the drift velocity). In the case of charged particles, both the pressure gradient and the friction force point upwards. In the case of neutrals, however, the later has the same value but points downward. The interplay of forces is such that both ions and neutrals experience the same acceleration regardless of their respective densities. It is worth to point out that, in their study on the support of neutrals against gravity in solar prominences, \citet{terradas2015} showed that ions remain static while neutrals slip down across the almost horizontal magnetic field with a small downward velocity. The drift velocities found by \citet{terradas2015} are of the order of a few m~s$^{-1}$, i.e., ten times larger than those obtained here, but rather small nonetheless. Analogous values to those in \citet{terradas2015} were found by \citet{gilbert2002} for the diffusion of neutral atoms in a hydrogen $+$ helium static prominence plasma.

The results obtained in Paper~I (for a fully ionized condensation) are also found here. In particular, the falling material displays two separate phases: first, the blob is accelerated and next it maintains a practically constant speed (Figures~\ref{fig_no_friction_vs_friction} and \ref{fig_ionization_degree_z_v}), with the duration of the first phase lasting longer for a denser blob. In addition, the emission of small-amplitude sound waves at the start of the descent is present in a partially ionized clump and makes itself visible by means of the oscillations in some curves of Figures~\ref{fig_blob_accel}, \ref{fig_ionization_degree_a}, and \ref{fig_velocity_drift2}. Finally, the important correlation between the blob maximum speed and the initial density ratio derived in Paper~I remains unchanged once partial ionization is included.

We have also assessed the importance of heating by collisions and the energy interchange between charged particles and neutrals, given by the terms $Q_i^{in}$ and $Q_n^{in}$ in Equations~(\ref{eqpie}) and (\ref{eqpn}). We have shown that an initial temperature imbalance between ions and neutrals in the blob is removed in a time of the order of 2~ms and that despite the brevity of this interval, it leads to unrealistic pressure gradients and velocity drifts in the blob edges. Nevertheless, these unrealistic features are innocuous because they do not affect the blob dynamics. We have repeated some numerical simulations with $Q_i^{in}=Q_n^{in}=0$ and have obtained identical results: the temporal variation of the blob height and velocity are the same, the accelerations caused by the forces in the momentum equation are the same, and the temporal variation of the drift velocity at the maximum density position is also unchanged. We conclude that the blob dynamics is insensitive both to the thermal structure of the blob and to the energy interplay between neutrals and ions, because both terms in the heat generation quantities $Q_i^{in}$ and $Q_n^{in}$ (see their definition in Equations~(\ref{eqqiin}) and (\ref{eqqnin})) are negligible. The first one, caused by heating because of friction between the two species, is proportional to $(\vi-\vn)^2$ and we have seen that the drift velocity is very small. Moreover, the second term, that accounts for the thermal energy transfer, is proportional to the temperature difference between the two species and, as mentioned before, $\Ti-\Tn$ vanishes very quickly and remains zero during the whole temporal evolution.

A final comment about the interaction between the charged and neutral portions of the condensation is needed. Here this interaction has been found to be strong enough to tie together ions and neutrals when the clump trajectory is vertical. Coronal rain blobs usually follow curved paths, so one may wonder if the friction force provided by ions would be sufficient to drag neutrals along the curved magnetic field, or else neutrals would slip through the field lines and would separate from the ions (contradicting the observations). This question cannot be answered with the one-dimensional model of this work and two-dimensional numerical simulations as in \citet{terradas2015} are required to ascertain whether the physics of a partially ionized blob would be qualitatively the same in the case of a trajectory guided by curved magnetic fields.

\section*{Acknowledgements}
J.T. acknowledges support from the Spanish Minis\-te\-rio de Educaci\'on y Ciencia through a Ram\'on y Cajal grant. R.S. acknowledges support from MINECO through a ``Juan de la Cierva'' grant, from MECD through project CEF11-0012, and from the ``Vicerectorat d'Investigaci\'o i Postgrau'' of the UIB. R.O., R.S., and J.T. acknowledge support from MINECO and FEDER funds through project AYA2014-54485-P. The work of T.Z. was supported by the Austrian Fonds zur F\"orderung der Wissenschaftlichen Forschung (FWF) project P26181-N27, by the European FP7-PEOPLE-2010-IRSES-269299 project-SOLSPANET, and by Shota Rustaveli National Science Foundation grant DI/14/6-310/12. The authors are also grateful to The Leverhulme Trust for funding under grant IN-2014-016 and thank the International Space Science Institute for the financial support and the facilities for two meetings on partially ionized plasmas in astrophysics and on coronal rain observations and modelling. The authors thank P. Antolin and D. Mart\'\i nez-G\'omez for useful comments that helped improve the manuscript. R.O. is indebted to D. W. Fanning for making available the Coyote Library of IDL programs (http://www.idlcoyote.com/).

%\bibliography{references}
%\bibliographystyle{apj}

\end{document}